\documentclass[useAMS, usenatbib]{mnras}
\usepackage{graphicx}
\usepackage{amssymb}
\usepackage{booktabs}
\usepackage{amsmath}
\usepackage{bm}
\usepackage{xcolor}
\usepackage{tikz}
\newcommand\ApJ{ApJ}
\newcommand\MNRAS{MNRAS}
\newcommand\AnA{A\&A}

\newcommand{\persec}{\,\mathrm{s}^{-1}}
\newcommand{\cc}{\,\mathrm{cm}^{-3}}

\newcommand{\cross}{\bm{\times}}
\newcommand{\J}{\bm{J}}
\newcommand{\B}{\bm{B}}
\newcommand{\E}{\bm{E}}

\newcommand{\sigmaH}{\sigma_\mathrm{H}}
\newcommand{\sigmaP}{\sigma_\mathrm{P}}
\newcommand{\sigmapara}{\sigma_\parallel}

\newcommand{\etaH}{\eta_\mathrm{H}}
\newcommand{\etaA}{\eta_\mathrm{A}}
\newcommand{\etaP}{\eta_\mathrm{P}}

\def\bq{\begin{equation}}
\def\eq{\end{equation}}
\def\alf{Alfv\'en\,}
\newcommand{\vv}{\bm{v}}
\def\va{v_A}
\def\Hp{\mbox{H}^+}
\def\Hep{\mbox{He}^+}
\def\cc{{\mbox{cm}^{-3}}}
\def\c1{{\mbox{cm}^{3}}}
\def\sigv{<\sigma\,v>}
\def\qq{\bm{q}}
\def\bets{\beta_i^*}
\def\nuis{\nu_i^*}
\def\nuns{\nu_n^*}
\def\xiis{\xi_i^*}
\def\xins{\xi_n^*}
\newcommand{\grad}{\nabla}
\newcommand{\divg}{\grad \cdot}
\newcommand{\curl}{\grad \cross}
\newcommand{\delt}[1]{\frac{\partial #1}{\partial t}}
\newcommand{\mH}{\mbox{H}}
\newcommand{\mHe}{\mbox{He}}
\newcommand{\tenq}[1]{\hbox{\oalign{$\bm{#1}$\crcr\hidewidth$\scriptscriptstyle\bm{\approx}$\hidewidth}}}
\def\bb{\bm{b}}
\title{Non-ideal Transport Processes in the Solar Atmosphere}
\author[B.P.Pandey and Mark Wardle]
        {B.P. Pandey and Mark Wardle \\
{School of Mathematical and Physical Sciences, Macquarie University, Sydney, NSW 2109, Australia}}
\date{\today}
\pagerange{\pageref{firstpage}--\pageref{lastpage}}
\pubyear{2025}
\begin{document}
\maketitle
\label{firstpage}
\begin{abstract}
Transport coefficients are calculated for a partially ionized plasma consisting of approximately $90\%$ hydrogen and $10\%$ helium, representative of a model solar atmosphere with an assumed magnetic field profile. The ion-Hall parameter, defined as the ratio of ion-cyclotron to ion collision frequency, is determined by considering dominant resonance charge exchange processes alongside less significant nonresonant ion-neutral collisions. Based on these calculations, we derive profiles for various transport coefficients. Our results demonstrate that thermal conductivity in partially ionized media—both parallel and perpendicular to the ambient magnetic field—is dominated by neutral particles. The perpendicular thermal conductivity components show weak dependence on the ion Hall parameter and remain comparable in magnitude to their parallel counterparts. Wave damping through neutral thermal conductivity may contribute significantly to solar atmospheric heating. These findings indicate that perpendicular thermal conductivity components are essential for accurate modelling of partially ionized regions, including photosphere-chromosphere transition layers, spicules, and coronal prominences.

\end{abstract}
\section{Introduction}
The solar atmosphere, like the solar nebula, consists of approximately $\sim 90\%$ hydrogen (H) and $\sim 10\%$ helium (He) by number, with a minor fraction ($0.1\%$) of heavier elements. Iron (Z=26) and other lighter metals have abundances roughly four orders of magnitude lower than hydrogen, while heavier elements such as nickel (Z=28) and copper (Z=29) are six to seven orders of magnitude less abundant \citep{F86}. 

For this study, we employ a simplified atmospheric model composed of hydrogen ($\sim 90\%$) and helium ($\sim 10\%$), representative of solar composition while omitting trace elements. Although hydrogen remains predominantly neutral near the temperature minimum region-where rare metals (Mg, Fe, C, Si, Al) constitute the primary ionic species-we assume that ionized hydrogen represents all ionic species throughout this region. This approximation, while offering considerable computational simplicity, does not significantly impact the accuracy of the calculated transport coefficients.

The ionization level in the solar atmosphere varies dramatically from the weakly ionized photosphere and lower chromosphere to the fully ionized corona, with the partially ionized middle and upper chromosphere positioned between these extremes. Additionally, relatively cool ($T\sim 10^4$\,\mbox{K}), dense ($\sim 10^{10}-10^{11}\,\cc$) large-scale ($5-10^2$ Mm) structures commonly observed as $\mbox{H}{\alpha}$-emitting plasma are embedded within the hotter ($T\sim 10^6\,\mbox{K}$), tenuous (density $\sim 10^{9}\,\cc$) X-ray emitting corona.

These structures maintain the same compositional profile as the broader solar atmosphere: approximately $90\%$ hydrogen and $10\%$ helium, consistent with solar and cosmic abundances. The term "prominence" describes a variety of such cool objects, which appear as dark filamentary features when viewed in $\mbox{H}{\alpha}$ absorption against the solar disc \citep{P24}. Due to their low temperatures, prominence plasmas are partially ionized, with electron-to-neutral hydrogen density ratios varying roughly between $0.1$ and $10$ \citep{Pa02}.

Solar structures including spicules, prominences, and atmospheric layers (photosphere, chromosphere, transition region) thus represent partially ionized systems comprising both plasma and neutral particles, with ionization fractions that vary systematically with altitude.

In the present work, we derive the parallel (with respect to the magnetic field), perpendicular, and cross thermal conductivities for the model solar atmosphere of \cite{F93} by considering dominant resonance charge exchange processes. Although viscosity coefficients in a partially ionized plasma have been given by \cite{PW22}, we recalculate these coefficients for the present model solar atmosphere in the presence of resonance charge exchange. We also derive expressions for the magnetic diffusion coefficients using the present model. The role of parallel and perpendicular conductivities in wave damping and chromospheric heating is examined and compared with other non-ideal heating mechanisms.

\section{Collisional Processes}

In a weakly ionized plasma, Coulomb collisions can be neglected, leaving only ion-neutral and electron-neutral collisions as significant processes. Partially ionized plasmas require consideration of all collision types: ion-ion, electron-electron, electron-ion, ion-neutral, and electron-neutral collisions. In fully ionized plasmas, plasma-neutral collisions become negligible, and the system is dominated by Coulomb interactions between charged particles.

The plasma is considered fully ionized in the collisional sense due to the long-range nature of Coulomb forces. However, Coulomb collisions differ fundamentally from neutral particle collisions—rather than discrete collision events, the long-range electromagnetic interactions result in continuous velocity randomization of charged particles \citep{BR65}.

\begin{figure}
\includegraphics[scale=0.30]{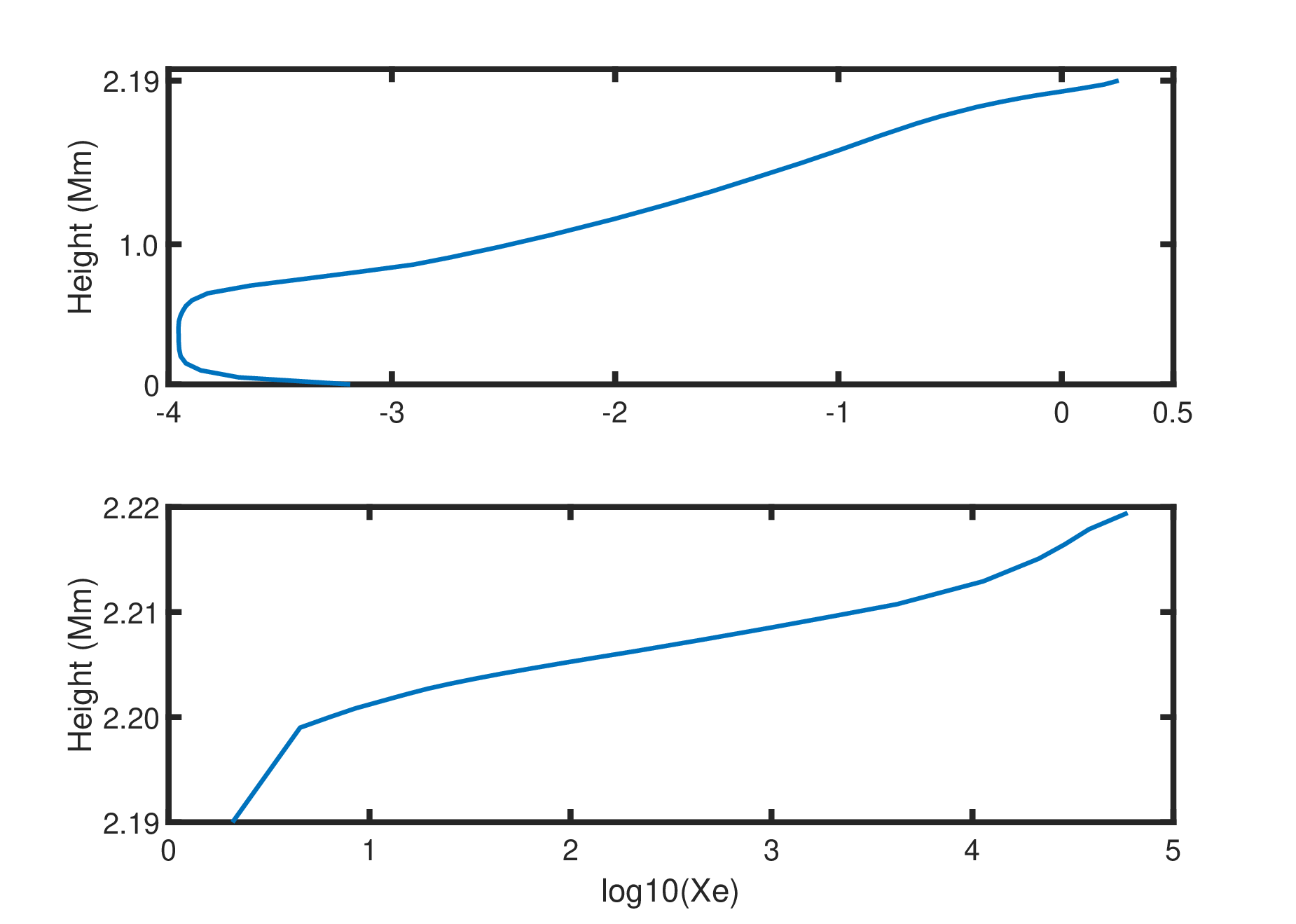}
\caption{The fractional ionization as a function of height is plotted for  model C of F93 in the figure above.}
\label{fig:F0}  
\end{figure}
Figure~\ref{fig:F0} demonstrates how the fractional ionization, defined as the ratio of electron ($n_e$) to neutral ($n_n$) number density,
\bq
Xe = \frac{n_e}{n_n}\,,
\eq
increases systematically with solar altitude. We adopted model atmosphere C of \cite{F93} for the above plot. This ratio rises from $10^{-4}$ in the photosphere to $10^{-2}$ above 1 Mm height in the middle chromosphere. The electron number density $n_e$ begins to exceed the neutral number density $n_n = n_H + n_{He}$ (where $n_H$ and $n_{He}$ represent the number densities of hydrogen and helium, respectively) only above approximately 2 Mm altitude.

In the transition region (lower panel) and beyond, plasma densities dominate over neutral densities by several orders of magnitude. This altitude-dependent ionization gradient fundamentally alters the collision dynamics: the plasma transitions from ion-neutral dominated interactions in the weakly ionized photosphere and lower chromosphere to ion-ion dominated interactions in the upper chromosphere and transition region.

The collision frequency $\nu_{j n}$ is expressed as
\bq
\nu_{j\,n} = \gamma_{j n}\,\rho_n  = \frac{\sigv_j}{m_n + m_j}\,\rho_n \,.
\label{eq:cfin}
\eq
Here $\sigv_j$ represents the rate coefficient for momentum transfer by collisions of the $j^{\mbox{th}}$ particle species. Further, $m_n$ represents the mean mass of neutrals, and $\rho_n=m_n\,n_n$ is the neutral mass density. 

For the neutral-neutral rate coefficients in cgs units \citep{GB86, F93} (hereafter F93)
\begin{eqnarray}
\sigv_{\mbox{H}-\mbox{H}}&=&2.26\times 10^{-9}\Big[1-\left(1.1-0.1\mbox{T}^{1/4}\right)^2\Big]\,,
\nonumber \\
\sigv_{\mbox{He}-\mbox{He}}&=&4\times 10^{-10}\,,
\nonumber \\
\sigv_{\mbox{He}-\mbox{H}}&=& 6.31\times 10^{-9}\,\mbox{T}^{-0.2}\,,
\end{eqnarray}
the collision frequencies becomes
\begin{eqnarray} 
\nu(\mbox{H},\mbox{H})&=&1.13\times 10^{-9}\Big[1-\left(1.1-0.1\mbox{T}^{1/4}\right)^2\Big]\frac{n(\mbox{H})}{\c1}\persec\,,
\nonumber\\
\nu(\mbox{He},\mbox{He})&=&4\times 10^{-10}\frac{n(\mbox{He})}{\c1}\persec\,,
\nonumber\\
\nu(\mbox{He},\mbox{H})&=&1.58\times10^{-9}\,\mbox{T}^{-0.2}\frac{n(\mbox{H})}{\c1}\persec\,.
\label{eq:nnc}
\end{eqnarray}

\begin{figure}
\includegraphics[scale=0.10]{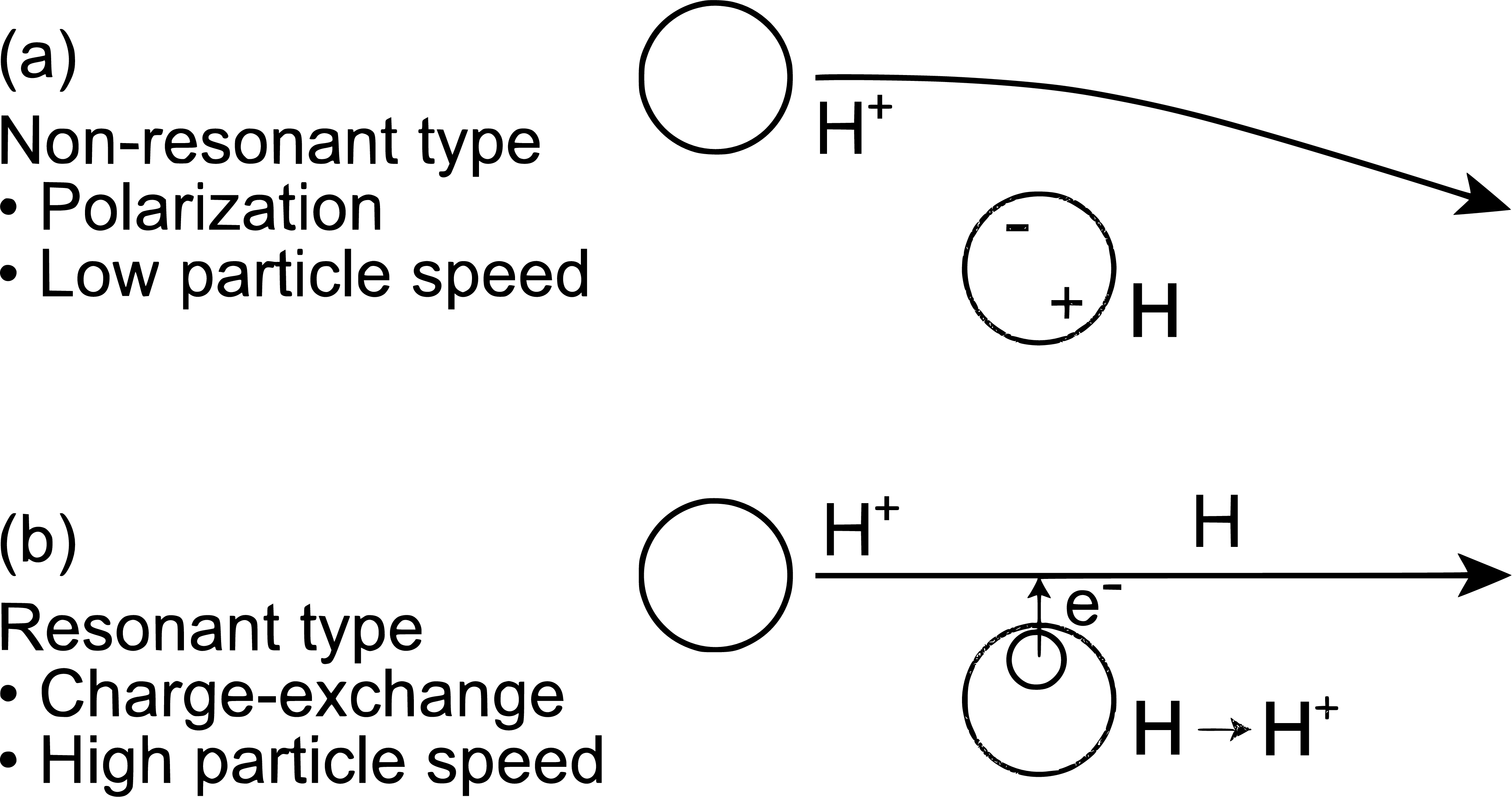}
\caption{Nonresonant and resonant interactions in the target rest frame for $\Hp-\mbox{H}$. (a) Nonresonant interaction arising from long-range polarization forces-distant collisions that dominant at low relative particle  speed (i.e., low temperatures); (b) resonant interaction resulting from electron transfer-close collisions that dominant at high speed (i.e., high temperatures).}
 \label{fig:RP}  
\end{figure}

Neutral-neutral ($\mbox{H}-\mbox{H}$ and $\mbox{He}-\mbox{He}$), ion-neutral ($\Hp-\mbox{H}\,,\Hp-\mbox{He}$ and $\Hep-\mbox{He}\,,\Hep-\mbox{H}$) and ion-ion ($\Hp-\Hp\,,\Hep-\Hep$) collisions dominate at different altitudes. The ion-neutral interaction can be resonant or non-resonant \citep{D58, D58a, O61, B66, BK73, S77}. Non-resonant interactions are possible for all ion-neutral pairs and arise from polarization attraction coupled with a short-range repulsion of neutral particles induced by approaching ions [Figure \ref{fig:RP}(a)]. Resonant collisions, on  the other hand, occur between ions and their parental neutrals, such as $\Hp$ and $\mbox{H}$, and result from electron exchange from the parental neutral to the ion [Figure \ref{fig:RP}(b)]. 

The nonresonant collision rate $\sigv$ for the short-range interactions (i.e., when the impact parameter is less than some critical value) is given by \citep{D11} 
\begin{eqnarray}
\sigv&=&2.41\,\pi\,Ze\,\left(
\frac{\alpha_N}{\mu_{in}}
\right)^{1/2}
\nonumber\\
&=&8.98\times 10^{-10}Z\left(
\frac{\alpha_N}{a_0^3}
\right)^{1/2}
\left(
\frac{m_H}{\mu_{in}}
\right)^{1/2}\mbox{cm}^3\persec\,.
\label{eq:cR}
\end{eqnarray}
Here 
\bq
\mu_{in}=\frac{m_i\,m_n}{m_i+m_n}\,,
\eq
is the reduced mass, where $m_i$ is the ion mass and $m_n$ is the neutral particle mass. Further, $\alpha_N$ is the polarizability of the neutral particle, $Z$ is the charge number of the ion, $e$ is the elementary charge, $a_0=5.292\times 10^{-9}\,\mbox{cm}$ is the Bohr radius, and $m_H$ is the hydrogen mass. The value of $\alpha_N/a_0^3=4.5$ and $1.383$  for $\mbox{H}$ \citep{C41, LL72} and $\mbox{He}$ \citep{TH72} respectively.

Accounting for both short- and long-range interactions gives the momentum transfer rate coefficient \citep{D11}
\bq
\sigv_{\mbox{mt}}=1.21\,\sigv\,.
\eq
Thus, from Eqs.~(\ref{eq:cfin}) and (\ref{eq:cR}) the nonresonant collision frequencies between $\Hp\,,\Hep$ and $\mbox{H}\,,\mbox{He}$ become
\begin{eqnarray}
\nu (\Hp\,,\mbox{H})&=&1.28\times 10^{-9}\quad \frac{n(\mbox{H})}{\c1}\persec\,,
\nonumber\\
\nu (\Hp\,,\mbox{He})&=&1.16\times 10^{-9}\quad \frac{n(\mbox{He})}{\c1}\persec\,,
\nonumber\\
\nu (\Hep\,,\mbox{He})&=&4.6\times 10^{-10}\quad\frac{n(\mbox{He})}{\c1}\persec\,,
\nonumber\\
\nu (\Hep\,,\mbox{H})&=&5.2\times 10^{-10}\quad \frac{n(\mbox{H})}{\c1}\persec\,.
\label{eq:NRCF}
\end{eqnarray}
As noted by \cite{O61} and \cite{D11}, the rate coefficient $\sigv$ in Eq.~(\ref{eq:cR}) is independent of temperature. Thus, nonresonant collision frequencies in Eq.~(\ref{eq:NRCF}) depend only on the neutral number density. With increasing temperature-and the corresponding increased mobility of ions and neutrals-the nature of collisions changes: charge exchange or resonant interactions between identical ion-neutral pairs become more frequent. 

Here we consider the following  resonant processes, 
\begin{eqnarray}
\Hp+\mbox{H}&\rightarrow& \mbox{H}+\Hp+\Delta\mbox{E}(=0)\,,\\
\nonumber
\Hep+\mbox{He}&\rightarrow&\mbox{He}+\Hep+\Delta\mbox{E}(=0)\,,
\end{eqnarray}
where $\Delta\mbox{E}$-the energy defect, is the difference between the internal
energy of the collision system before and after the collision. The resonant charge exchange (RCE) cross-section is relatively large, resulting in significant momentum transfer during RCE encounters. It has been shown
theoretically  \citep{D58a, BK73, SN09} that the RCE cross-section peaks at zero
energy and decrease with increasing collision energy $E$ as
\bq
\sigma=\left(a-b\,ln E\right)^2\,,
\label{eq:cx}
\eq
where $a$ and $b$ (in cm) are constants that are different for different gases. 

We note that in several recent works \citep{PB19, G24} express the RCE cross-section as a linear function of energy. While this linear form matches the Barnett atomic data to within $10\%$ for relative ion-neutral speeds  between $4.8\times 10^5\,\mbox{cm}/{s}$ and $1.4\times 10^8\,\mbox{cm}/{s}$ or equivalently between $0.12\,\mbox{eV}$ and $10\,\mbox{keV}$ for hydrogen \citep{MS12}, the quadratic form, Eq.~(\ref{eq:cx}) has strong experimental foundation.

Cross section fits to nearly $30$ years of experimental data for $\Hp-\mbox{H}$ collisions [Fig.~1a] and the corresponding parameterized formula in Table 1 of \cite{LS05} suggest that the theoretical cross-section formula, Eq.~(\ref{eq:cx}) is accurate to within approximately $\pm 10\%$. Therefore, we use Eq.~(\ref{eq:cx}) for calculating the collision frequency. The resonant $\Hp-\mbox{H}$ and $\Hep-\mbox{He}$  collision frequencies, valid for $T>50\,\mbox{K}$,  are \citep{SN09}
\begin{eqnarray}
\frac{\nu^R (\Hp\,,\mbox{H})}{2.65\times 10^{-10}}&=&\frac{n(\mbox{H})}{\c1}\,\mbox{T}^{1/2}\,\left(1-0.083\,\log_{10} \mbox{T}\right)^2 \persec\,,
\nonumber\\
\frac{\nu^R (\Hep\,,\mbox{He})}{8.73\times 10^{-11}}&=&\frac{n(\mbox{He})}{\c1}\,\mbox{T}^{1/2}\,\left(1-0.093\,\log_{10}\mbox{T}\right)^2
\persec\,.
\nonumber\\
&&
\label{eq:RC1}
\end{eqnarray}
\begin{figure}
\includegraphics[scale=0.30]{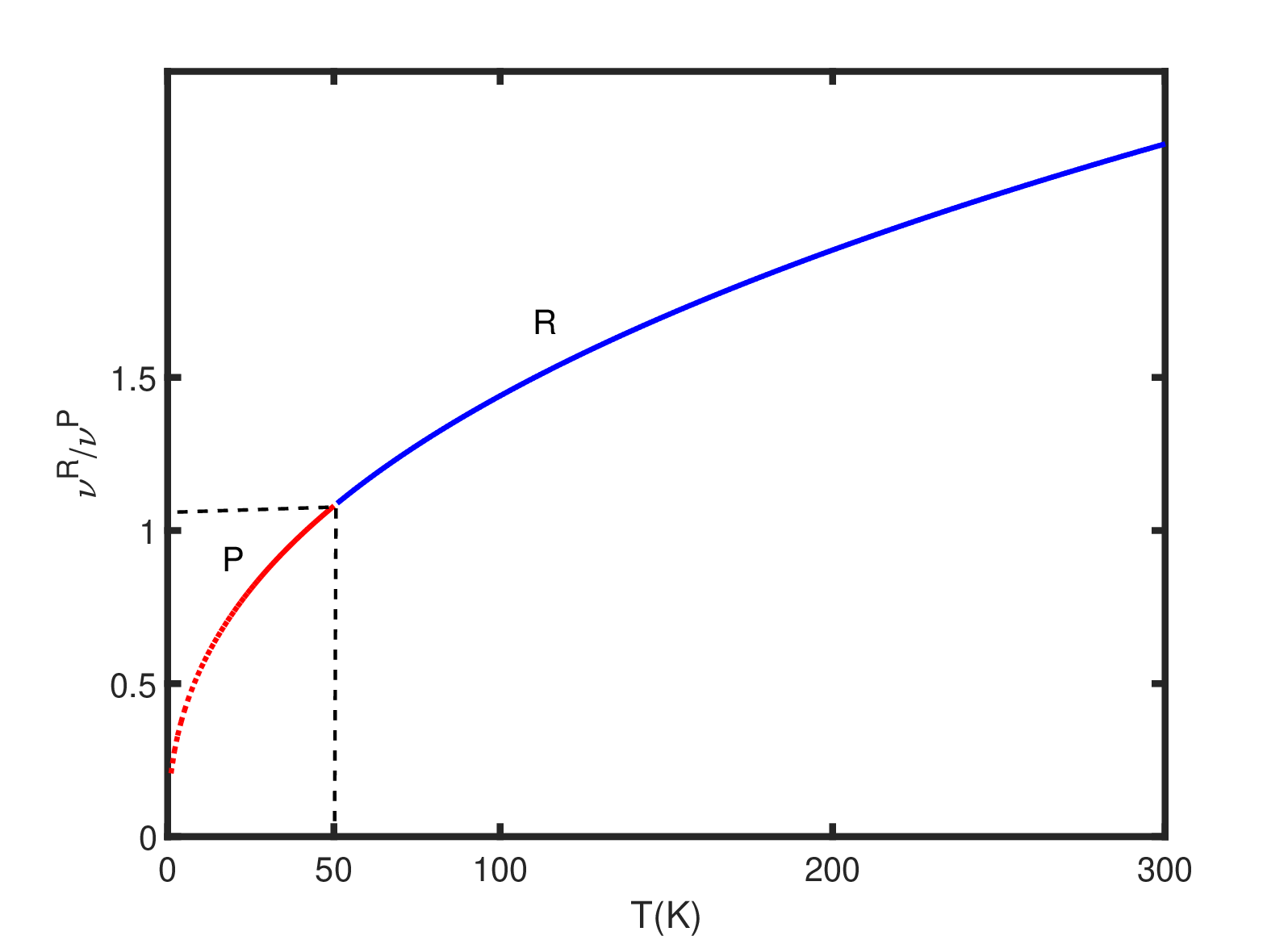}
\caption{Ratio of collision frequencies as a function of temperature for hydrogen atoms is shown in the figure above. Labels R and P corresponds to resonant (R) 
and polarization/nonresonant (P) frequencies, respectively.}
 \label{fig:FRP}  
\end{figure}
Note that there is a significant difference between the nonresonant ($\nu^P$) and resonant ($\nu^R$) collision frequencies. Whereas nonresonant frequencies $\nu^P$ depend only on the neutral number density, the resonant frequencies $\nu^R$ additionally depend on temperature as $\sqrt{T}$. This temperature dependence is physically intuitive, since charge exchange relies on the random thermal motion of particles; higher random speed leads to more frequent exchanges of identity between ions and neutrals, represented as $\mathrm{X}^+ \rightleftarrows \mathrm{X}$.

The ratio of resonant to nonresonant collision frequencies for hydrogen atoms is plotted against temperature in Fig.~(\ref{fig:FRP}). The inclusion of helium does not affect this relationship. We observe that nonresonant collisions (dotted curve) dominate at lower temperatures. For hydrogen, resonant interactions become significant only above $50\,\mathrm{K}$. Between $50$-$300\,\mathrm{K}$, both types of interactions are comparable in magnitude. At higher temperatures ($T>300\,\mathrm{K}$), ions and their parent neutral particles interact predominantly via resonant charge exchange (RCE). In this process, as an ion and neutral approach each other, an electron is transferred from the neutral to the ion, effectively exchanging their identities [Figure~\ref{fig:RP}(b)]; a fast-moving ion becomes a fast neutral after collision, resulting in significant momentum transfer between the colliding particles. It is important to note that polarization interactions between $\mathrm{H}$ and $\mathrm{H}^+$ cannot be entirely neglected, even at temperatures as high as $10^4\,\mathrm{K}$ \citep{GB86}. For instance, at typical prominence temperatures (approximately $7000\,\mathrm{K}$), polarization interactions contribute around $15\%$ to the total momentum transfer cross-section, with resonant charge transfer accounting for the remaining $85\%$ \citep{GHH02}.

\begin{figure}
\includegraphics[scale=0.30]{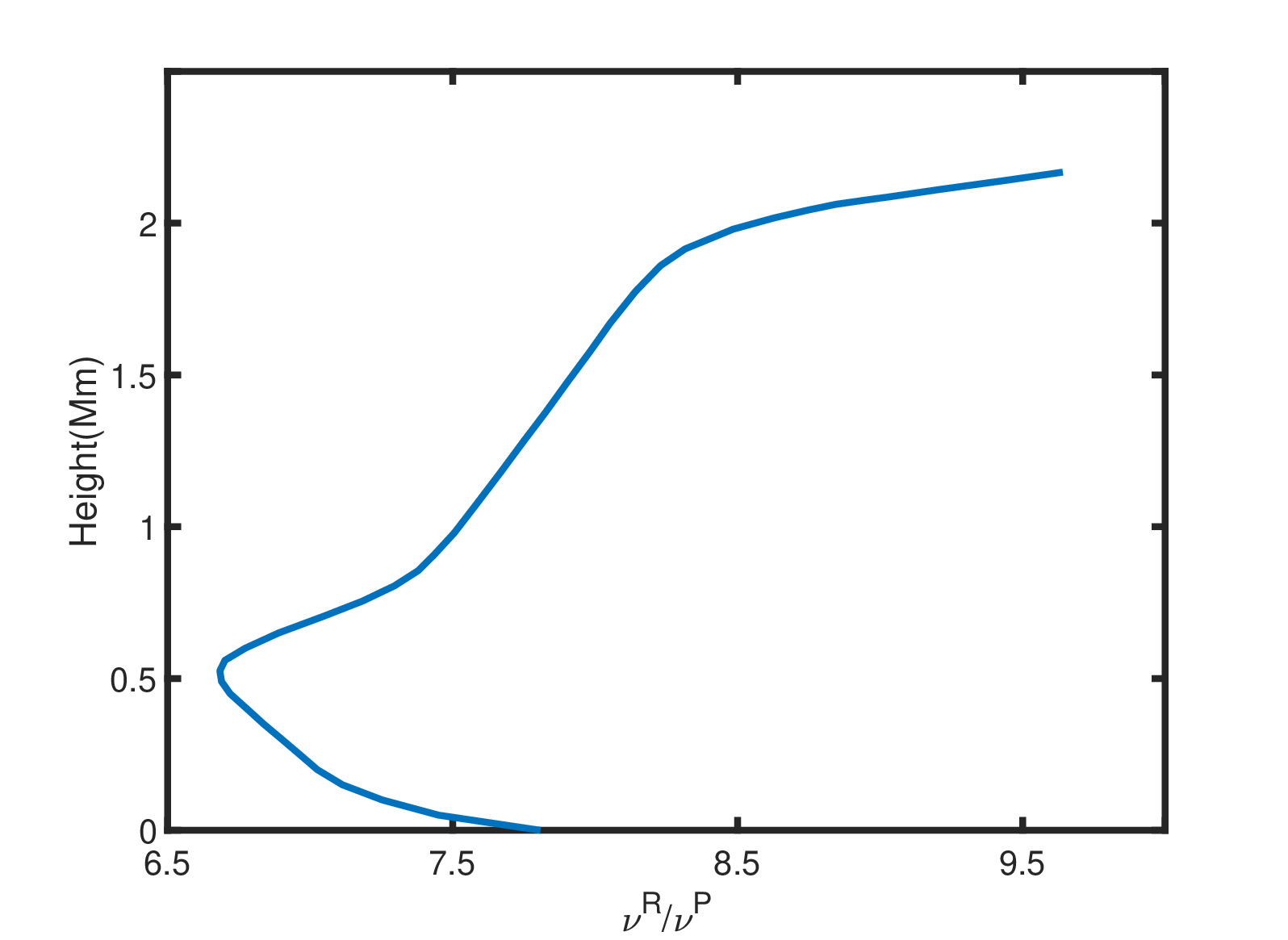}
\caption{Height variation of the collision frequency ratio for hydrogen atoms in model C of F93.}
 \label{fig:FRP1}  
\end{figure} 
In Fig~(\ref{fig:FRP1}) we plot the $\nu^R$ vs. $\nu^P$ ratio for the solar atmosphere using model C of F93.  Clearly, the resonant interaction is approximately an order of magnitude larger than the nonresonant interaction.  Therefore, the ion Hall beta, which measures ion magnetization, will be an order of magnitude smaller than that calculated using nonresonant collision frequencies.      

Adopting the magnetic field profile 
\begin{equation}
B = B_0\,\left(\frac{n_n}{n_0}\right)^{0.3}\,,
\label{eq:scl}
\end{equation}
where subscript $0$ corresponds to the reference value at the footpoint, 
we calculate the ion and electron Hall parameters 
\begin{equation}
\beta_j=\frac{\omega_{cj}}{\nu_{j}}\,,
\label{eq:IhB}
\end{equation}  
where $j=e,i$ and the cyclotron frequency $\omega_{cj}$ is given by 
\begin{equation}
\omega_{cj}=\frac{|q_j|\,B}{m_j\,c}\,.
\label{eq:cycf}
\end{equation}
Here $q_j$, $B$, $m_j$, and $c$ represent the charge, magnetic field, mass of the plasma particles, and the speed of light, respectively. The symbol $\nu_j$ represents the total plasma ($\mathrm{H}^+$, $\mathrm{He}^+$, $\mathrm{e}^-$)-neutral ($\mathrm{H}$ and $\mathrm{He}$) collision frequency
\begin{equation}
\nu_j=\sum_n\nu_{jn}\,.
\label{eq:sum_n}
\end{equation}
Furthermore, for hydrogen ($\mathrm{H}^+-\mathrm{H}$) and helium ($\mathrm{He}^+-\mathrm{He}$)
pairs, the collision frequency is the sum of polarization (P) and resonant (R) components:  
\begin{equation}
\nu_{jn}= \nu_{jn}^P + \nu^R_{jn}\,.
\end{equation}

The electron-hydrogen collision frequency is given by \citep{D83, PG08}
\bq
\nu(e\,,\mH)=8.28\times 10^{-10}\quad \mbox{n(H)}\quad \mbox{T}_e^{1/2}\,,
\label{eq:cen1}
\eq
where $n(H)$ is the hydrogen number density and $T_e$ is the electron temperature. 

The electron-helium collision frequency is given by \citep{PG08, SN09}
\bq
\nu(e\,,\mHe)=
4.6\times10^{-10}\quad\mbox{n(He)}\quad\mbox{T}_e^{1/2}\,,
\eq
where $n(\text{He})$ is the helium number density.

The ion-Hall parameter, $\beta_i$ [Eq.~(\ref{eq:IhB})], consists of two components: (a) a temperature-independent polarization part, $\beta_i^{\text{P}}$, and (b) a temperature-dependent resonant part (charge exchange), $\beta_i^{\text{R}}$. Therefore, both the polarization and resonant parts contribute to the ion magnetization.

The polarization part of the ion-Hall parameter for H$^+$ is given by
\bq
\beta_i^{\text{P}} = 9.6 \times 10^{-2} \left(\frac{B}{\rho_H}\right) \,, 
\label{eq:Ibet1}
\eq
which is independent of temperature and varies as $B/\rho_H$. Assuming $n_H/n_0 = \rho_H/\rho_0$, where $\rho_0$ is the reference value at the footpoint, and using Eq.~(\ref{eq:scl}), the above equation becomes
\bq
\beta_i^{\text{P}} = 9.6 \times 10^{-2} \left(\frac{B_0}{\rho_0}\right) \left(\frac{\rho_0}{\rho_H}\right)^{0.7} \,. 
\label{eq:Ibet2}
\eq
Thus, the polarization part of the ion-Hall parameter $\beta_i^{\text{P}}$ varies inversely as $\rho_H^{-0.7}$ for known values of magnetic field and density near the footpoint.

The resonant part of the ion-Hall parameter $\beta_i^{\text{R}}$ is given by
\bq \beta_i^{\text{R}} = 6 \times 10^{-11} \left(\frac{B}{\rho_H}\right) T^{-1/2} \left(1 - 0.083 \log_{10} T\right)^{-2} \,. 
\label{eq:res1}
\eq
Clearly, the resonant component of the ion-Hall parameter, in addition to its dependence on $B/\rho_H$, is proportional to the inverse square root of temperature, apart from the weak temperature-dependent correction factor $\left(1 - 0.083 \log_{10} T\right)^{-2}$ in Eq.~(\ref{eq:res1}).

The electron-Hall parameter is given by
\bq
\beta_e = 3.3\times 10^{-8} \left(\frac{B}{\rho_H}\right)\,\mbox{T}_e^{-1/2}
\eq
and has a similar dependence to $\beta_i^{\text{R}}$ on both $B/\rho_H$ and $T_e^{-1/2}$.

\subsection{Ion-magnetization}

The transport properties of a magnetized plasma depend, in addition to its thermodynamic state, on the mobility of plasma particles—namely, on the electron and ion  magnetization parameters, $\beta_e$ and $\beta_i$ respectively. As noted above, in the weakly ionized mixture of hydrogen and helium, transport processes are dominated by ion-neutral  [Eq.~(\ref{eq:sum_n})] and electron-neutral [Eq.~(\ref{eq:cen1})] collisions.

When the ion-Hall parameter $\beta_i < 1$, ions are deflected by neutrals each time they attempt to complete a helical orbit around the magnetic field lines. They must restart on a new helical path only to be deflected again, resulting in a random walk of ions across the magnetic field. In this regime, ions do not follow the magnetic field and are therefore unmagnetized. The resulting transverse drift (with respect to the ambient magnetic field) between unmagnetized ions and magnetized electrons ($\beta_e > 1$) gives rise to the Hall effect \citep{PW06,PW08}. 

Collision frequencies between neutrals and charged particles dominate the ion-cyclotron frequency in the photosphere and lower chromosphere, where $\beta_i < 1$. However, with increasing altitude, the ion-neutral collision frequency decreases more rapidly than the ion-cyclotron frequency, and ions become magnetized ($\beta_i > 1$). Thus, with increasing altitude, the relative slippage of magnetized ions against the sea of neutrals causes ambipolar diffusion in the middle and upper chromosphere. Note that in the $\beta_i > 1$ regime, finite Larmor radius effects (which induce viscous momentum transport via neutrals) compete with ambipolar diffusion of the magnetic field \citep{PW22, PW24}.

With increasing altitude, the plasma becomes partially ionized in the upper chromosphere-transition region, where frequent ion-ion and ion-electron collisions may deflect ions from their helical orbits and cause random walk behavior. As a result, at higher altitudes, ion magnetization becomes a function of ion-ion and ion-electron collisions in addition to ion-neutral collisions. 

Therefore, the definition of the ion-Hall parameter $\beta_j$ given by Eq.~(\ref{eq:IhB}) should be generalized to reflect the partially ionized nature of the plasma. The collision frequency $\nu_j$ given by Eq.~(\ref{eq:nnc}) and the neutral collision frequency given by Eq.~(\ref{eq:sum_n}) should become \citep{Z02}
\bq
\nu_j = 0.3 \, \nu_{jj} + \sum_{\substack{k \neq j}} f_{jk} \, \nu_{jk} \,, 
\label{eq:cfZ}\eq
where $j = i, n$ and
\bq
f_{jk} = \frac{m_j + 0.6 \, m_k \, A_{jk}^*}{m_j + m_k} \,. 
\label{eq:wtZ}
\eq
The above expression for $f_{jk}$ differs from that given by \citep{Z02} owing to our slightly different definition of the momentum transfer collision frequency. Here, $A_{jk}^*$ is the ratio of the Chapman-Cowling integrals, which for the hard-sphere collision model is unity: $A_{jk}^* = 1$.

We rewrite Eq.~(\ref{eq:cfZ}) separately for the ion and neutral collision frequencies:
\begin{align}
\nu_i &= 0.3 \, \nu_{ii} + \sum_{n} f_{in} \, \nu_{in} + f_{ie} \, \nu_{ie} \,, 
\label{eq:cfi} \\
\nu_n &= 0.3 \, \nu_{nn} + \sum_{i} f_{ni} \, \nu_{ni} + f_{ne} \, \nu_{ne} \,. 
\label{eq:cfx}
\end{align}
The electron collision frequency is \citep{Z02}
\bq
\nu_e = 0.3 \, \nu_{ee} + 0.6 \, \nu_{ei} + 0.6 \, \nu_{en} \,. 
\label{eq:ef}
\eq

From the expression for the electron-ion collision frequency $\nu_{ei}$ \citep{PW08}:
\bq
\nu_{ei}=51\,n_i\,T^{-3/2}\,\text{s}^{-1}\,,
\eq
we obtain the ion-electron collision frequency using $\nu_{ie} = \left(m_e/m_i\right)\,\nu_{ei}$. For hydrogen and helium ions, this yields:
\begin{align}
\nu(\text{H}^+,e) &= 2.78\times 10^{-2}\,n(\text{H}^+)\,T^{-3/2}\,\text{s}^{-1}\,, \\
\nu(\text{He}^+,e) &= 6.9\times 10^{-3}\,n(\text{He}^+)\,T^{-3/2}\,\text{s}^{-1}\,.
\end{align}

The ion-ion collision frequency, $\nu_{ii}= \sqrt{\frac{2\,m_e}{m_i}}\,\nu_{ei}$, becomes:
\begin{align}
\nu(\text{H}^+,\text{H}^+) &= 9.2\times 10^{-4}\,n(\text{H}^+)\,T^{-3/2}\,\text{s}^{-1}\,, \\
\nu(\text{He}^+,\text{He}^+) &= 1.9\times 10^{-4}\,n(\text{He}^+)\,T^{-3/2}\,\text{s}^{-1}\,.
\end{align}

For collisions between different ion species \citep{SN09}:
\begin{align}
\nu(\text{H}^+,\text{He}^+) &= 1.14\,n(\text{He}^+)\,T^{-3/2}\,\text{s}^{-1}\,, \\
\nu(\text{He}^+,\text{H}^+) &= 0.28\,n(\text{H}^+)\,T^{-3/2}\,\text{s}^{-1}\,.
\end{align}

Denoting the composite ion collision terms: 
\begin{eqnarray} 
C(\Hp)=0.68\,\Big[\nu(\Hp\,,\mHe) + \nu(\Hp\,,\Hep)\Big] 
\nonumber\\
+ 0.8\,\nu(\Hp\,,\mH) \,,
\nonumber \\
C(\Hep)=\Big[0.92\,\nu(\Hep\,,\mH)+\nu(\Hep\,,\Hp)\Big]
\nonumber\\
+0.8\,\nu(\Hep\,,\mHe)\,,
\end{eqnarray}
the ion collision frequency, Eq.~(\ref{eq:cfi}), becomes:
\begin{eqnarray}
\nu(\Hp)&=&0.3\,\nu(\Hp\,,\Hp)  + \nu(\Hp\,,e) + C(\Hp)\,,
\nonumber\\
\nu(\Hep)&=&0.3\,\nu(\Hep\,,\Hep) +  \nu(\Hep\,,e) + C(\Hep)\,.
\nonumber\\
\label{eq:cfx1}
\end{eqnarray}
Similarly, defining the neutral collision terms:
\begin{eqnarray} 
D(\mH)&=&0.68\,\Big[\nu(\mH\,,\Hep)+\nu(\mH\,,\mHe)\Big]+ 0.8\,\nu(\mH\,,\Hp)\,,
\nonumber \\
D(\mHe)&=&0.92\,\Big[\nu(\mHe\,,\Hp)+\nu(\mHe\,,\mH)\Big]+0.8\,\nu(\mHe\,,\Hep)\,,
\nonumber\\
\end{eqnarray}
the neutral collision frequency, Eq.~(\ref{eq:cfx}) becomes:
\begin{eqnarray}
\nu(\mH)&=&0.3\,\nu(\mH\,,\mH) + \nu(\mH\,,e) + D(\mH)\,,
\nonumber\\
\nu(\mHe)&=&0.3\,\nu(\mHe\,,\mHe) +\nu(\mHe\,,e) + D(\mHe)\,.
\label{eq:cfx2}
\end{eqnarray}

For heat flux calculations, the collision frequency $\nu_{j}$ [Eq.~(\ref{eq:cfZ})] is replaced by the modified collision frequency $\nu_j^*$, which accounts for the enhanced momentum transfer in heat conduction \citep{Z02}:
\begin{equation}
\nu_j^*=0.4\,\nu_{j\,j}+\sum_{k\neq j} \left(1.6+3\,\frac{m_j}{m_k}+ 1.3\,\frac{m_k}{m_j}\right) h_{j\,k}\,\nu_{j\,k}\,,
\label{eq:nus}
\end{equation}
where the weighting factor is defined as
\begin{equation}
h_{j\,k}=\frac{\mu_{j\,k}}{m_{j}+m_{k}}\,,
\end{equation}
and $\mu_{j\,k}$ is the reduced mass of the colliding pair.

The expression above assumes a hard-sphere molecular model. For the hydrogen-helium gas mixture, Eq.~(\ref{eq:nus}) simplifies to
\begin{align}
\nu(\Hp)^*&=0.4\,\nu(\Hp\,,\Hp)  + 3\,\nu(\Hp\,,e) + C^*(\Hp)\,, \label{eq:cfxs1a}\\
\nu(\Hep)^*&=0.4\,\nu(\Hep\,,\Hep) +  3\,\nu(\Hep\,,e) + C^*(\Hep)\,, \label{eq:cfxs1b}
\end{align}
where 
\begin{align}
C^*(\Hp) &= 1.21\,\Big[\nu(\Hp\,,\mHe)+\nu(\Hp\,,\Hep)\Big] + 1.47\,\nu(\Hp\,,\mH)\,, \\
C^*(\Hep) &= 0.16\,\Big[\nu(\Hep\,,\Hp)+\nu(\Hep\,,\mH)\Big] + 0.25\,\nu(\Hep\,,\mHe)\,.
\end{align}

For the neutral species, the modified collision frequencies following Eq.~(\ref{eq:nus}) are
\begin{align}
\nu(\mH)^*&=0.4\,\nu(\mH\,,\mH) + 3\,\nu(\mH\,,e) + D^*(\mH)\,, \label{eq:cfx2a}\\
\nu(\mHe)^*&=0.4\,\nu(\mHe\,,\mHe) + 3\,\nu(\mHe\,,e) + D^*(\mHe)\,, \label{eq:cfx2b}
\end{align}
where 
\begin{align} 
D^*(\mH) &= 1.21\,\Big[\nu(\mH\,,\Hep)+\nu(\mH\,,\mHe)\Big] + 1.47\,\nu(\mH\,,\Hp)\,, \\
D^*(\mHe) &= 2.23\,\Big[\nu(\mHe\,,\Hp)+\nu(\mHe\,,\mH)\Big] + 1.47\,\nu(\mHe\,,\Hep)\,.
\end{align}

We distinguish between different formulations of the ion-Hall parameter [Eq.~(\ref{eq:IhB})] based on the collision frequency used: $\beta_{i,\nu}$ employs the standard collision frequency from Eq.~(\ref{eq:cfi}), while $\beta_{i,\chi}$ uses the modified collision frequency from Eq.~(\ref{eq:nus}). This distinction is physically motivated, as the viscosity and thermal diffusivity coefficients in partially ionized plasmas depend on $\beta_{i,\nu}$ and $\beta_{i,\chi}$, respectively. Both parameters represent the sum of contributions from the two ion species: $\beta_{i,\nu}=\beta_{\Hp,\nu}+\beta_{\Hep,\nu}$ and $\beta_{i,\chi}=\beta_{\Hp,\chi}+\beta_{\Hep,\chi}$.

\begin{figure}
\includegraphics[scale=0.30]{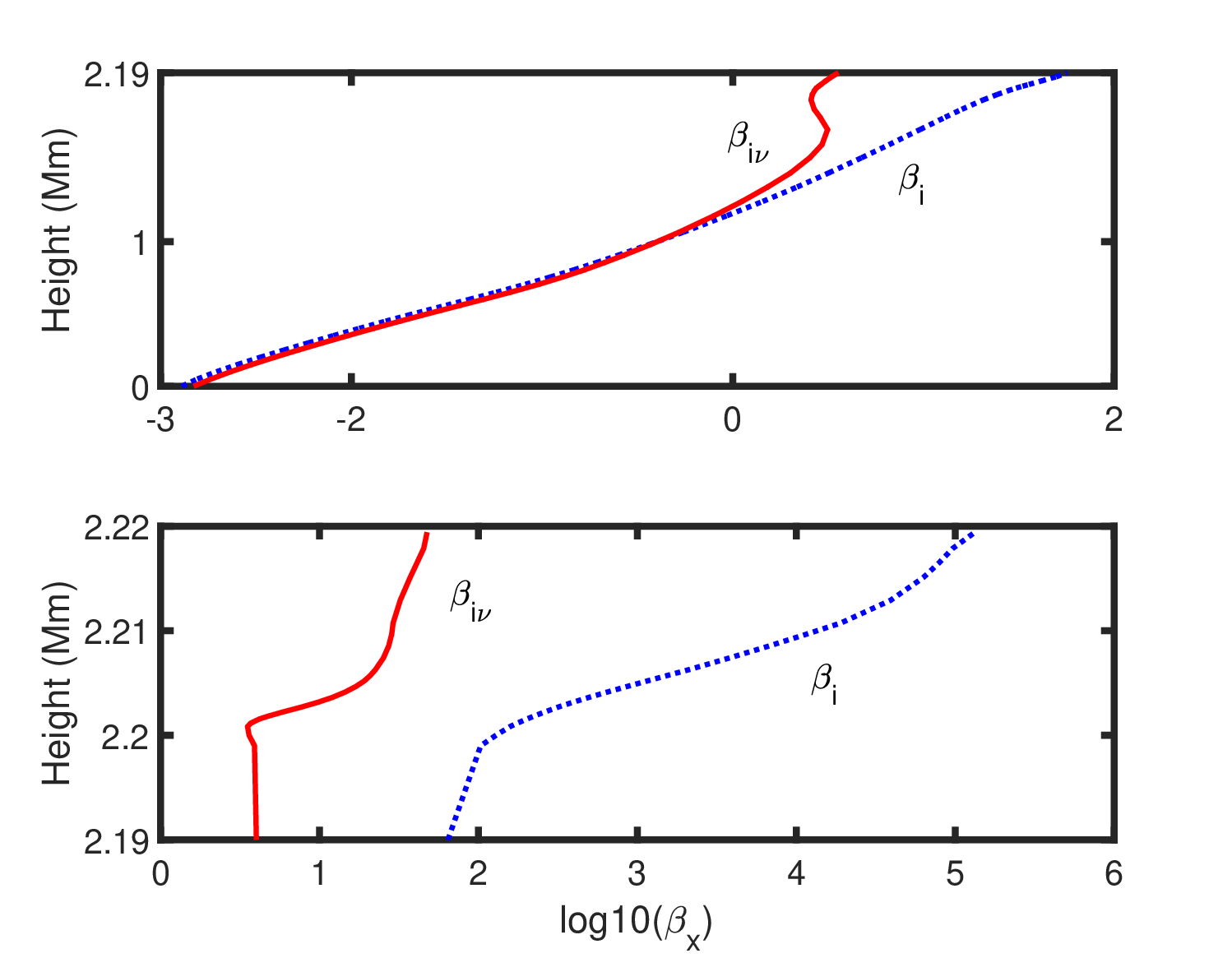}
\caption{The height variation of the ion-Hall ($\beta_{i}$) and viscous-Hall ($\beta_{i,\nu}$) parameters are displayed in the above panels.}
 \label{fig:F_beta}  
\end{figure}

Figure~(\ref{fig:F_beta}) reveals that the ion-Hall parameters $\beta_i$ (appropriate for weakly ionized media) and $\beta_{i,\nu}$ (for partially ionized plasmas) are nearly identical in the photosphere and lower chromosphere, consistent with the weakly ionized nature of these regions. However, as ionization increases with height, $\beta_{i}$ exceeds $\beta_{i,\nu}$ due to two factors: the declining neutral density and the growing importance of ion-ion interactions. This divergence becomes pronounced in the transition region (lower panel), where the difference spans several orders of magnitude. Since $\beta_{i,\chi}$ exhibits a profile similar to $\beta_{i,\nu}$, it is omitted from the figure for the sake of clarity.

\section{Plasma Conductivity/ Magnetic Diffusivity}
The electrical conductivity of a weakly ionized gas is determined by calculating the drift velocities of charged particles in response to an applied electric field $\mathbf{E}'$ in the neutral reference frame, then summing the contributions from all charged species to obtain the total current density \citep{C76}. For each charged species $i$ with mass $m_i$ and charge $Z_i e$, the drift velocity $\mathbf{v}_i$ relative to the neutrals (characterized by mean mass $m_n$ and density $\rho_n$) is found by balancing three forces: the electric force from the applied field, the magnetic force due to particle motion in the magnetic field, and the collisional drag from interactions with neutrals:
\begin{equation}
    n_i\,Z_i\,e \left( \E' + \frac{\vv_i}{c} \cross\B\right)=\rho_i\,\nu_{in}\,\vv_i\,.
    \label{eq:vj}
\end{equation}
Inverting eq (\ref{eq:vj}) for $\vv_i$, and forming the current density $\J=\sum_i n_i\,e\,Z_i\,\vv_i$ yields
\begin{equation}
    \J = \sigmapara \E'_{\parallel} + \sigmaH \bb\cross\E'_\perp + \sigmaP 
    \E'_\perp
    \label{eq:J-sigmaE}
\end{equation}
where $\bb = \B /B$, and $\E'_\parallel$ and $\E'_\perp$ are the components of $\E'$ parallel and perpendicular to $\B$, respectively. Ohm, Hall, and Pedersen conductivities are given by \citep{C76, WN99}:
\begin{eqnarray}
    \sigma_O &=& \frac{e\,c}{B}\sum_j n_j\,|Z_j|\,\beta_j\,,
    \label{eq:sigmaO} \\
    \sigma_H &=& -\frac{e\,c}{B}\sum_j \frac{n_j\,Z_j\,\beta_j^2 }{ 1+\beta_j^2}
    = \frac{ec}{B}\sum_i \frac{n_j\,Z_j}{ 1+\beta_j^2}\,,
    \label{eq:sigmaH} \\
    \sigma_P &=& \frac{e\,c}{B}\sum_j \frac{n_j\,|Z_j|\,\beta_j}{ 1+\beta_j^2}\,,
    \label{eq:sigmaP} \\
	\sigma_\perp &=& \sqrt{\sigma_H^2 + \sigma_P^2}\,,
\end{eqnarray}
where we have used the charge neutrality condition $\sum_j n_j\,Z_j = 0$ in deriving the second form of $\sigma_H$ in Eq.~(\ref{eq:sigmaH}). Here $e$ is the elementary charge.

When magnetic fluctuations are negligible, such as in high-frequency, short-wavelength electrostatic instabilities like the Farley-Buneman mode, the induction equation becomes redundant \citep{G14}. However, for magnetic fluctuations, the induction equation plays a crucial role and can be derived by inverting Eq.~(\ref{eq:J-sigmaE}) to solve for $\mathbf{E}'$:

\bq
\E'=\frac{\J}{\sigma_O}-\left(\frac{\sigma_P}{\sigma_\perp^2}-\frac{1}{\sigma_O}\right) \J_\perp-\frac{\sigma_H}{\sigma_\perp^2}\J\cross\bb\,.
\eq
Thus, the induction equation becomes
\begin{eqnarray}
\delt \B = \curl\left[
\left(\vv\cross\B\right) - \frac{4\,\pi\,\eta_O}
{c}\,\J - \frac{4\,\pi\,\eta_H}{c}\,\J\cross\bb
\right. \nonumber\\
\left.
+ \frac{4\,\pi\eta_A}{c}\,
\left(\J\cross\bb\right)\cross\bb
\right]\,,
\label{eq:ind}
\end{eqnarray}
where Ohm ($\eta_O$), Hall ($\eta_H$) and Pedersen ($\eta_P$) diffusivities are 
\begin{eqnarray}
    \eta_O &=& \frac{c^2}{4\pi\sigma_O}\,,\\[6pt]
    \label{eq:etaO}
    \etaH  &=& \frac{c^2}{4\pi\sigma_\perp}\,\frac{\sigma_H}{\sigma_\perp}\,,\\[6pt]
    \label{eq:etaH}
    \etaP &=& \frac{c^2}{4\pi\sigma_\perp}\,\frac{\sigma_P}{\sigma_\perp}\,, \\[6pt]
    \label{eq:etaH}
    \eta_\perp &=& \sqrt{\eta_P^2 + \eta_H^2} \quad=\quad \frac{c^2}{4\pi\sigma_\perp}\,, \\[6pt]
    \label{eqn:etaperp}
    \etaA &=& \etaP-\eta_O\,,
    \label{eq:etaA}
\end{eqnarray}
and we have also tacked on the perpendicular ($\eta_\perp$) and ambipolar ($\etaA$) resistivities.

\begin{figure}
\includegraphics[scale=0.30]{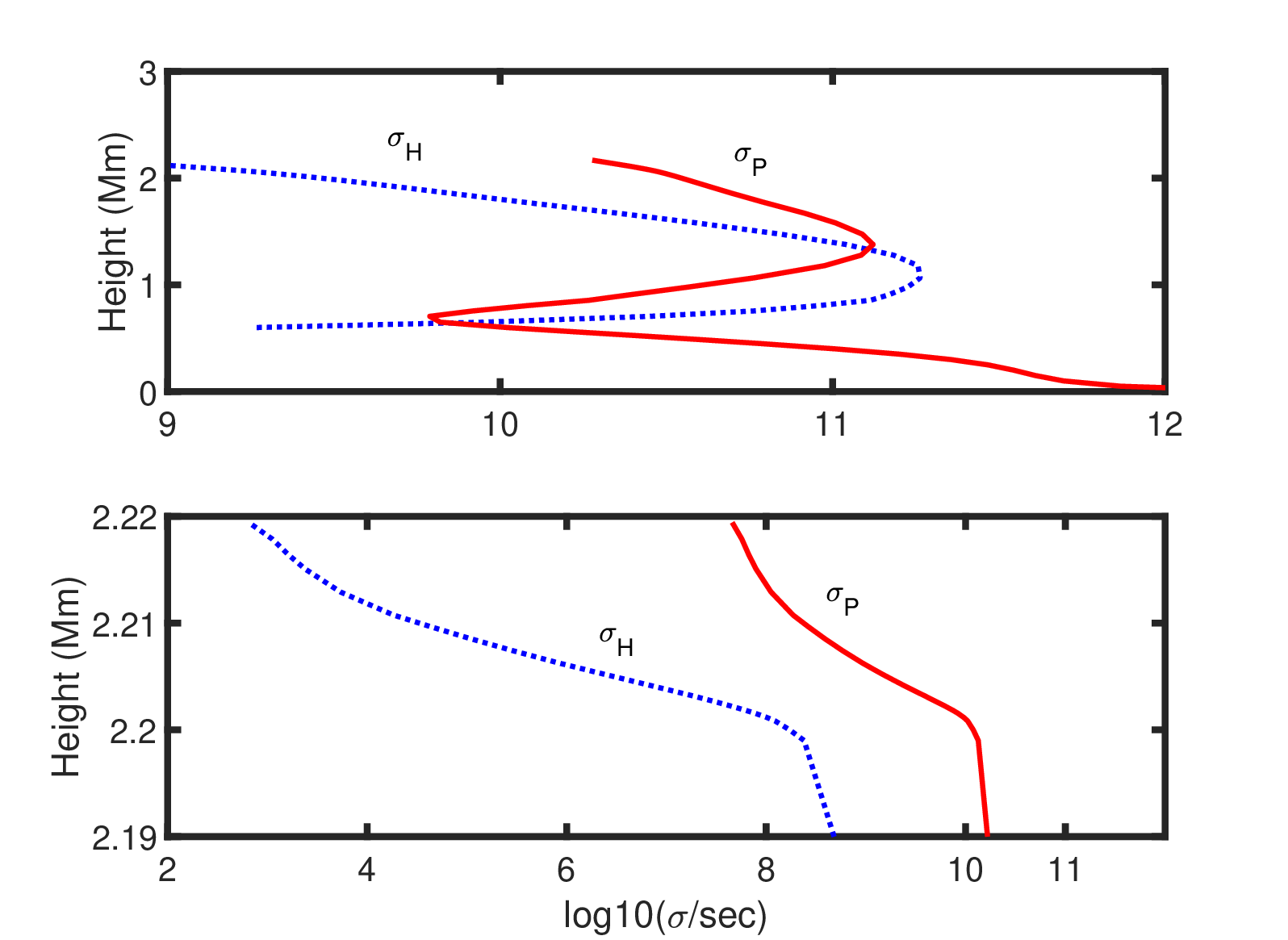}
\caption{The Hall ($\sigma_H$) and Pedersen ($\sigma_P$) conductivities are plotted as functions of height using atmospheric model C from \citet{F93}. The magnetic field profile employed in the calculations is described in the paper, with a base field strength of $B_0=100\,\text{G}$ at the photospheric footpoint.}
 \label{fig:Fcond}  
\end{figure} 
\begin{figure}
\includegraphics[scale=0.30]{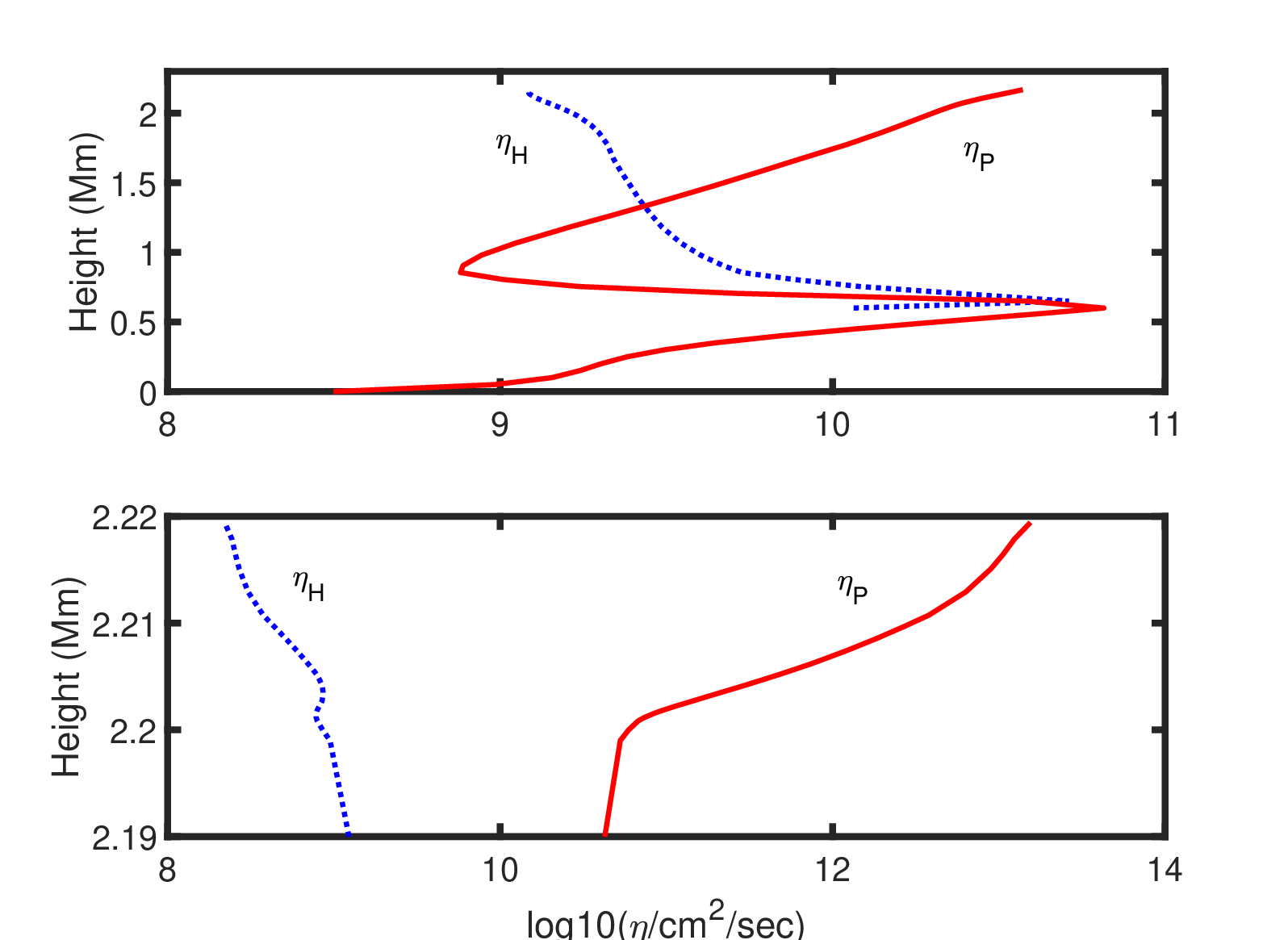}
\caption{The Hall ($\eta_H$) and Pedersen ($\eta_P$) magnetic diffusivities are plotted as functions of height using the same atmospheric model and magnetic field configuration as described in the previous figure.}
 \label{fig:Fdiff}  
\end{figure} 

In the limit $\beta_e\gg\beta_i$, the Hall and ambipolar diffusivities can be expressed as
\begin{align}
\eta_H &= \frac{v_A^2}{\omega_H} \propto X_e^{-1}\left(\frac{B}{n_n}\right)\,\text{cm}^2\,\text{s}^{-1}\,, \label{eq:etaH_approx}\\
\eta_A &= \frac{v_A^2}{\nu_{ni}}\propto X_e^{-1}\left(\frac{B}{n_n}\right)^2\,\text{cm}^2\,\text{s}^{-1}\,, \label{eq:etaA_approx}
\end{align}
where $\omega_H=\frac{\rho_i}{\rho_n}\omega_{ci}$ is the Hall frequency and $v_A=B/\sqrt{4\pi\rho}$ is the \alf velocity. The parallel diffusivity in terms of fractional ionization is \citep{BG09}
\bq
\eta_O = 230\,T^{1/2}\,X_e^{-1}\,\text{cm}^2\,\text{s}^{-1}\,.
\label{eq:eta_parallel}
\eq

The characteristic length scales over which Hall and ambipolar diffusion operate are
\begin{align}
L_H &= \frac{v_A}{\omega_H}\propto X_e^{-1}\,n_n^{-1/2}\,, \label{eq:LH}\\
L_A&= \frac{v_A}{\nu_{ni}}\propto X_e^{-1}\,n_n^{-3/2}\,B\,. \label{eq:LA}
\end{align}

Notably, the Hall scale is independent of the magnetic field strength. This reflects the underlying physics: the Hall effect arises from the relative transverse drift between unmagnetized ions ($\beta_i<1$) and magnetized electrons ($\beta_e>1$) \citep{PW06, PW08}. The Hall length $L_H$ therefore represents the scale over which ions, due to frequent collisions with neutrals, cannot follow the magnetic field lines and thus exhibit field-independent behavior. In essence, the Hall scale characterizes the length scale of ion demagnetization. As expected, the Hall effect diminishes with increasing ion magnetization.

In contrast, the ambipolar scale $L_A$ exhibits a linear dependence on magnetic field strength ($L_A \propto B$), while the 
Ohm scale, $L_{\eta}$ varies inversely with field strength ($L_{\eta} \propto B^{-1}$).

Figure~(\ref{fig:Fcond}) shows the Hall and Pedersen conductivities from Eqs.~(\ref{eq:sigmaH})--(\ref{eq:sigmaP}), with the corresponding magnetic diffusivities presented in Fig.~(\ref{fig:Fdiff}). Consistent with previous studies \citep{PW12, PW13, K14}, we observe a clear stratification of diffusion mechanisms with height. Ohm diffusion dominates in the photosphere, where $\eta_P \approx \eta_O$, while ambipolar diffusion becomes dominant in the upper chromosphere, where $\eta_P \approx \eta_A$. Hall diffusion $\eta_H$, which exhibits intermediate values between Ohm and ambipolar diffusivities, plays a crucial role in the lower and middle chromosphere.

Magnetic diffusivities appear to be similar in both quiet and active regions of the Sun [e.g., Fig.~5 of \cite{K14}]. However, these analytical studies fail to capture the strong variations observed in the chromosphere, as revealed by numerical simulations \citep{M12}. Additionally, the commonly used expressions for the collision frequencies $\nu(\mathrm{H}, \mathrm{H}^+)$ and $\nu(\mathrm{He}, \mathrm{H}^+)$ exhibit a $T^{1/2}$ temperature dependence, similar to that of the RCE collision frequency [Eq.~(\ref{eq:RC1})] \citep{K14, M12, PW22}. These temperature-dependent forms do not account for polarization effects, which, due to the $\sigma \propto 1/v$ dependence, result in a temperature-independent value of $\sigv = \text{const}$.

\section{Viscosity Coefficients}
 Recall that both neutral and plasma particles participate in momentum transport through parallel, perpendicular, and gyro viscosities. The viscous stress tensor can be decomposed into three distinct components according to their orientation relative to the magnetic field \citep{BR65, Z02}:
\bq
\tenq{\Pi} = \tenq{\Pi}_{\parallel} +  \tenq{\Pi}_{\perp} + \tenq{\Pi}_{\Lambda}\,.
\eq
Here $\parallel\,,\perp\,,\Lambda$ are the parallel [$\bb(\bb\cdot\nabla)$], perpendicular [$-\bf{b}\cross(\bf{b}\cross \nabla$)] and cross ($\bf{b}\cross \nabla$) terms with respect to the magnetic field direction $\bb=\B /B$. This decomposition captures the fundamental anisotropy imposed by the magnetic field on transport processes. The parallel component $\tenq{\Pi}_{\parallel}$ describes transport along field lines where particles move freely, while the perpendicular component $\tenq{\Pi}_{\perp}$ and gyro component $\tenq{\Pi}_{\Lambda}$ represent the more restricted cross-field transport mechanisms.

Each constituent of the partially ionized plasma—electrons, ions, and neutrals—contributes to the viscosity coefficients. However, under the assumption of thermal equilibrium among all species, the electron viscosity contribution becomes negligible and can be safely omitted from the analysis \citep{Z02, PW22}.
  
In terms of the rate of strain tensor $\tenq{W}$, the stress tensor $\tenq{\Pi}$ is given by
\bq
\tenq{\Pi} = - \eta_0\,\tenq{W}_{0} - \eta_1\,\tenq{W}_{1} - \eta_2\,\tenq{W}_{2} + \eta_3\,\tenq{W}_{3} + \eta_4\,\tenq{W}_{4}\,,
\label{eq:vstx}
\eq
where the first term represents the parallel stress component, the second and third terms correspond to perpendicular stress components, and the final two terms represent cross stresses, commonly referred to as gyroviscous stresses.

The various ion and neutral $\eta$ coefficients depend on  plasma pressure, magnetization  and collision frequencies. Tensor $\tenq{W}_k$ in the above Eq.~(\ref{eq:vstx}) is expressed in terms of strain rate tensor $\tenq{W}:$ 
\bq
\tenq{W} = \grad\vv + \left(\grad\vv\right)^{T} - \frac{2}{3}\,\tenq{I}\divg\vv\,,
\label{eq:stn}
\eq
which decomposes into five distinct tensor components: 
\begin{eqnarray}
\tenq{W}_{0}&=&  \frac{3}{2}\,\left(\bb\cdot\tenq{W}\cdot\bb\right)\left(\bb\bb-\frac{1}{3}\tenq{I}\right)\,,
\nonumber\\
\tenq{W}_{1}&=&\tenq{I}_{\perp}\cdot\tenq{W}\cdot \tenq{I}_{\perp}
-\frac{1}{2}\tenq{I}_{\perp} \tenq{I}_{\perp}\bm{:}\tenq{W}\,,
\nonumber\\
\tenq{W}_{2}&=&\tenq{I}_{\perp}\cdot\tenq{W}\cdot \bb\bb + \bb\bb \cdot \tenq{W}\cdot  \tenq{I}_{\perp}\,,
\nonumber\\
\tenq{W}_{3}&=&\frac{1}{2}\left(\bb\cross\tenq{W}\cdot\tenq{I}_{\perp} - \tenq{I}_{\perp}\cdot \tenq{W}\cross\bb \right)\,,
\nonumber\\
\tenq{W}_{4}&=& \bb\cross\tenq{W}\cdot\bb\bb - \bb\bb\cdot\tenq{W}\cross \bb\,.
\end{eqnarray}
Here, $\tenq{I}$ denotes the identity tensor and $\tenq{I}_{\perp}=\tenq{I}-\bb\bb$ is the projection tensor onto the plane perpendicular to the magnetic field direction $\bb$. 

Assuming $T_i = T_e= T_n = T$, the ion and neutral viscosity coefficients $\eta$ are \citep{Z02, V14}
\begin{eqnarray}
\eta_{i\,0} = \frac{p_i}{2\,\nu_i}\,\xi_i\,\Delta_{\eta}^{-1}\,,\quad
\eta_{i\,1} = \frac{\eta_{i\,0}}{1+\Delta_{\eta}^{-2}\,\beta_i^2}\,,
\nonumber\\  
\eta_{i\,2} = \eta_{i\,1}\Big[\frac{\beta_i}{2}\Big]\,,
\quad
\eta_{i\,3} = \eta_{i\,1}\,\beta_i\,\Delta_{\eta}^{-1}\,,
\label{eq:ivis}
\end{eqnarray}
and
\begin{eqnarray}
\eta_{n\,0} = \frac{p_n}{2\,\nu_n}\,\xi_n\,\Delta_{\eta}^{-1}\,,\quad
\eta_{n\,1} =\eta_{n\,0}\, \frac{1+\beta_i^2\,\xi_n^{-1}\,\Delta_{\eta}^{-1}
}{1+\beta_i^2\,\Delta_{\eta}^{-2}}\,,
\nonumber\\  
\eta_{n\,2} = \eta_{n\,1}\Big[\frac{\beta_i}{2}\Big]\,,
\quad
\eta_{n\,3}=\eta_{n\,0}\,\,\beta_i\,\Delta_{\eta}^{-1}\,\frac{1-\left(\frac{\Delta_{\eta}}{\xi_n}\right)}{1+\left(\frac{\beta_i^2}{
\Delta_{\eta}^2}\right)}\,,
\label{eq:nvis}
\end{eqnarray}
\bq
\eta_{i\,,n\,4} = \eta_{i\,,n\,3}\Big[\frac{\beta_i}{2}\Big]
\eq
where $\beta_i\equiv \beta_{i\,\nu}$ and the square bracket $[\beta_i/2]$ means that wherever $\beta_i$ occurs they should be replaced  by $\beta_i/2$ to get the new viscosity coefficients.  Further, $p_i=n_i\,k_B\,T$ and $p_n=n_n\,k_B\,T$ are the ion and neutral kinetic pressure respectively and $k_B$ is Boltzmann constant.

Defining
\bq
g_{\alpha\,\beta}=\frac{m_{\alpha}\,m_{\beta}}
{\left(m_{\alpha}+m_{\beta}\right)^2}\,\Big[1-0.6\left(\frac{m_\beta}{m_\alpha}\right)\,A_{\alpha\,\beta}^{*}
\Big]\,,
\eq
we may write other parameters in the viscosity coefficients as 
\begin{eqnarray}
\xi_{\alpha}&=& 1+g_{\alpha\,\beta}\,\left(\frac{\nu_{\alpha\,\beta}}{\nu_{\beta}}\right)\,,
\nonumber\\
\Delta_{\eta}&=&1-g_{\alpha\,\beta}^2\,\frac{\nu_{\alpha\,\beta}\nu_{\beta\,\alpha}}{\nu_{\alpha}\nu_{\beta}}\,.
\end{eqnarray}

Note that $\xi_n \sim \xi_i\sim \Delta_{\eta} \sim 1$. Further, for hard sphere collision model $A_{i\,n}^{*} =1$. Thus the ion and neutral viscosity coefficients, Eqs.~(\ref{eq:ivis})-(\ref{eq:nvis}) becomes
\bq
\eta_{i\,0} \approx \frac{p_i}{2\,\nu_i}\,,\quad
\eta_{n\,0} \approx \frac{p_n}{2\,\nu_n}\,.
\eq
We emphasize that the quantities $\nu_i$ and $\nu_n$ in the above expressions are defined by Eqs.~(\ref{eq:cfi}) and (\ref{eq:cfx}), respectively. Although many recent studies \citep{S15, PB19, L24} assume $\nu_i = \nu_{ii}$ and $\nu_n = \nu_{nn}$, this approximation is not consistent with the partially ionized nature of the solar plasma. As pointed out by \citet{K04}, at minimum one should consider $\nu_i \rightarrow \nu_{ii} + \nu_{in}$ and $\nu_n \rightarrow \nu_{nn} + \nu_{ni}$. The exact expressions, Eqs.~(\ref{eq:cfi})–(\ref{eq:cfx}), assign different numerical weights to the various collision processes.

It is important to note that ion–neutral collisional drift gives rise to ambipolar diffusion of the magnetic field. Consequently, neglecting $\nu_{in}$ in the viscosity formulation is inconsistent with the presence of ambipolar drift in the medium and leads to an overestimation of the ion viscosity. Similarly, neglecting $\nu_{ni}$ results in an overestimation of the neutral viscosity.

\begin{figure}
\includegraphics[scale=0.25]{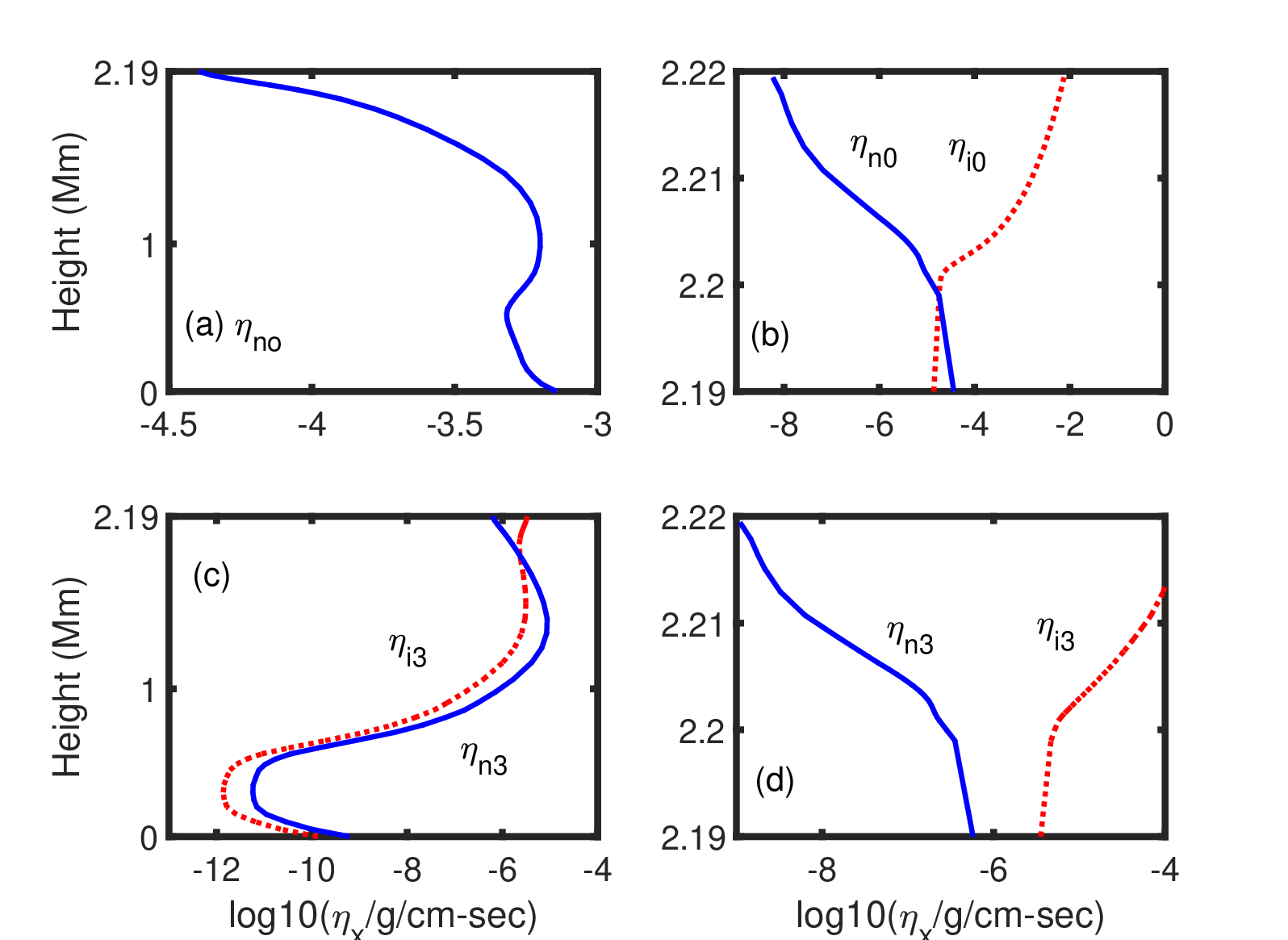}
\caption{Panels (a) and (b) show the parallel neutral viscosity coefficient $\eta_{n0}$ as a function of height. Panel (b) additionally displays the ion parallel viscosity coefficient $\eta_{i0}$ (dotted line). Panels (c) and (d) present the gyroviscosity coefficients for neutrals and ions versus height.
}
\label{fig:F9}  
\end{figure}

Note that for the ion viscosity components, $\tenq{\Pi}_{\perp}$/$\tenq{\Pi}_{\parallel}\sim \eta_{i1}/\eta_{i0}\sim1/\beta_i^2\,,$  $\tenq{\Pi}_{\Lambda}$/$\tenq{\Pi}_{\parallel}\sim \eta_{i3}/\eta_{i0} \sim 1/\beta_i$. Therefore, we would expect that parallel ion viscosity is the main contributor to the viscous momentum transport for the ion fluid, as the perpendicular ion viscosity is only important when the ions are weakly magnetized i.e. $\beta_i\rightarrow 1$ or $k\,R_L\sim \mathcal{O}(1)$ \citep{PW22}. Here $R_L=c_s/\omega_{ci}$ is the ion-Larmor radius and $k\sim \nu_i/c_s$ is the wavevector.  In the $\beta_i\rightarrow 1$ limit however, both the Hall and gyroviscous effects are equally important and kinetic theory is the 
proper framework to investigate the momentum transport in  the plasma.

Due to very weak dependence of the perpendicular component of neutral viscosity on the ion-Hall parameter, $\beta_i$, the ratio of neutral viscosity components, $\tenq{\Pi}_{\perp}$/$\tenq{\Pi}_{\parallel}\sim \eta_{n1}/\eta_{n0}\sim \mathcal{O}(1)\,$. Therefore, both the parallel and perpendicular component of the neutral viscosity contribute to the viscous momentum transport in equal measure. The weak dependence of the perpendicular component of the neutral viscosity on the $\beta_i$ becomes important in the presence of shear flow and is at the heart of viscous shear instability \citep{PW22, PW24}. 

In Fig.~(\ref{fig:F9}) we show the parallel [panels (a) and (b)] and gyroviscous [panels (c) and (d)] coefficients for the hydrogen. The helium contribution to the viscosity coefficients, compared to the hydrogen is orders of magnitude smaller. In the top left panel (a) only parallel coefficient of the neutral viscosity is shown. The perpendicular coefficients is similar to the parallel component. Further, we do not show in this frame the parallel ion viscosity coefficient which is orders of magnitude smaller than the neutral component. Clearly parallel viscosity coefficient due to neutral hydrogen dominates the  ionized hydrogen viscosity coefficient in  the photosphere-chromosphere. In the top right panel (b) corresponding to the transition region the ionized hydrogen contribution to the parallel viscosity coefficient (after $2.2\,\mbox{Mm}$) dominates the neutral viscosity coefficient. To summarize, viscosity in the partially ionized solar atmosphere is: (i) the sum of parallel and perpendicular neutral viscosity in the photosphere-lower transition region and (ii) only parallel ion viscosity closer to the upper transition region. The gyroviscosity coefficients due to neutral and ionized hydrogen [panels (c) and (d)] are comparable in the photosphere-chromosphere and are orders of magnitude smaller than the parallel and perpendicular components [panels (a) and (b)].  With increasing altitude, ionized component of gyroviscosity is the main contributor to the total gyroviscosity [panel (c)]. 

\begin{figure}
\includegraphics[scale=0.25]{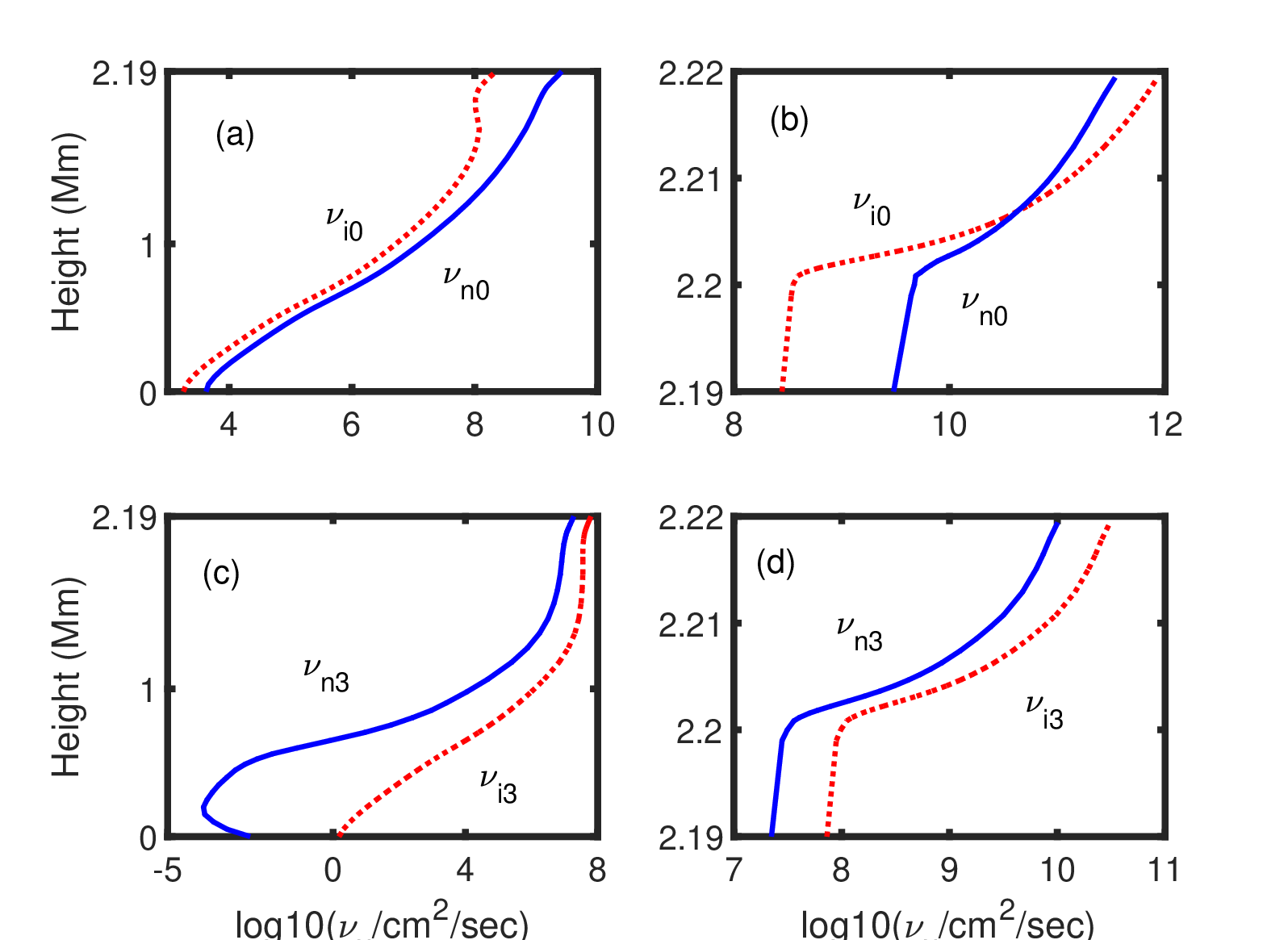}
\caption{Panels (a) and (b) display the parallel neutral and ion viscosities $\nu_{n0}$ and $\nu_{i0}$ versus height, while panels (c) and (d) show the corresponding gyroviscosities  $\nu_{n3}$ and $\nu_{i3}$ versus height.
}
 \label{fig:F10}  
\end{figure}

In Fig.~(\ref{fig:F10}) the parallel and gyroviscosities are plotted against height for the ion and neutral viscosities. Here the viscosities are sum of both the hydrogen and helium components, i.e. 
\bq
\nu_{j\,k}=\frac{\eta_{j k\,\mH}}{\rho_H} + \frac{\eta_{j k\,\mHe}}{\rho_{He}}\,.
\eq
Like the viscosity  coefficient profiles in Fig.~(\ref{fig:F9}), it is clear from the top panels (a)-(b) in  Fig.~(\ref{fig:F10}) that the parallel (and perpendicular) component of the neutral viscosity is the main contributor to the total viscosity in the photosphere-lower transition region. It is only closer to the coronal transition layer that the parallel ion viscosity becomes larger than the neutral counterpart. However, as can be seen from the lower (c) and (d) panels of the figure, the ion gyroviscosity is the main contributor to the total gyroviscosity in the entire  photosphere-chromosphere-transition region. Thus, we conclude that both the parallel and perpendicular viscous momentum transport in the photosphere-chromosphere and lower transition region is mainly due to neutral viscosity. It is only closer to the corona that viscous momentum transport is parallel to the magnetic field and is due to ions. The perpendicular momentum transport is negligible $\sim \mathcal{O}(1/\beta_i^2$) in this region.

\begin{figure}
\includegraphics[scale=0.25]{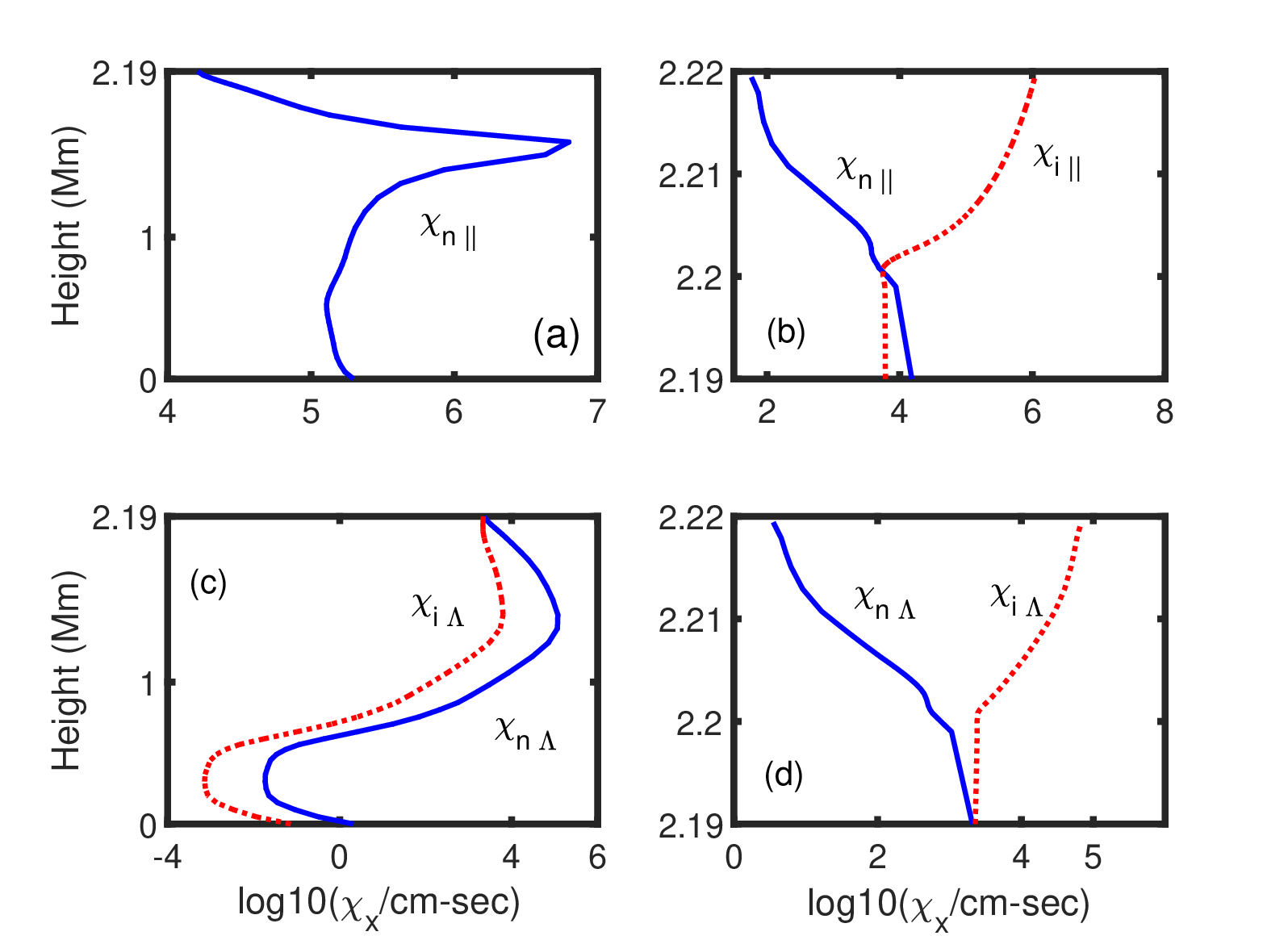}
\caption{Panel (a) shows the parallel thermal conductivity coefficient $\chi_{n ||}$ for neutral particles. Panel (b) displays both parallel neutral and ion thermal conductivity coefficients versus height. Panels (c) and (d) present the neutral and ion cross conductivity coefficients versus height.
}
 \label{fig:F5}  
\end{figure}

\section{Thermal Conductivity}
In a fully ionized plasma, the perpendicular and cross-thermal conductivities are reduced relative to the parallel conductivity by factors of $1/\beta_i^2$ and $1/\beta_i$, respectively \citep{BR65}. Since the thermal conductivity determines the direction of heat flux, and the parallel component greatly exceeds the perpendicular component in a fully ionized case, magnetic field lines may become visible in the corona through emission at appropriate temperatures \citep{WS21}.

In partially ionized plasma, however, thermal conduction involves contributions from both ions and neutrals, but the neutral contribution in the solar photosphere-chromosphere exceeds the ion contribution by orders of magnitude. Therefore, the assumption that ion thermal conductivity alone adequately describes thermal conduction throughout the solar atmosphere \citep{N22} is not justified. Moreover, as demonstrated below, the perpendicular component of neutral conductivity has the same order of magnitude as the parallel component. This occurs because, like perpendicular neutral viscosity, perpendicular neutral conductivity depends only weakly on the ion-Hall parameter $\beta_i$. Consequently, numerical models that consider only field-aligned thermal conductivity \citep{C12, B21} provide an inadequate description of chromospheric dynamics.

The heat flux $\qq$ is \citep{BR65, Z02}
\bq
\qq=-\chi_{\parallel}\grad_{\parallel} T - \chi_{\perp}\grad_{\perp} T - \chi_{\Lambda} \grad T\cross\bb\,.
\eq
Here assuming $T_i = T_e= T_n = T$, the thermal conductivity coefficients for ions become 
\begin{eqnarray}
\chi_{i\parallel} = \frac{5}{2}\frac{k_B}{m_i}\frac{p_i}{\nuis}\,\xiis\,\Delta_{\lambda}^{-1}\,,\quad
\chi_{i\perp} = \frac{\chi_{i\parallel}}{1+\Delta_{\lambda}^{-2}\,{\bets}^2}\,,
\nonumber\\  
\chi_{i\Lambda}=\bets\,\Delta_{\lambda}^{-1}\,\chi_{i\perp}\,,
\label{eq:tci}
\end{eqnarray}
and
\begin{eqnarray}
\chi_{n\parallel}&=&\frac{5}{2}\frac{k_B}{m_n}\frac{p_n}{\nuns}\,\xins\,{\Delta_{\lambda}}^{-1}\,,
\nonumber \\
\chi_{n\perp}&=&\chi_{n\parallel}
\frac{1+\Delta_{\lambda}^{-1}\left(\xins\right)^{-1}
{\bets}^2}{1+\Delta_{\lambda}^{-2}\,{\bets}^2}\,,
\nonumber\\
\chi_{n\Lambda}&=&\bets\,\Delta_{\lambda}^{-1}\,\chi_{n\parallel}
\frac{1-\Delta_{\lambda}^{-1}\,\left(\xins\right)^{-1}
}{1+\Delta_{\lambda}^{-2}\,{\bets}^2}\,,
\label{eq:tcn}
\end{eqnarray}
for neutrals. Here
\bq
\xiis =1-b_{n\,i}\,,\quad \xins =1-b_{i\,n}\,,\quad \Delta_{\lambda} = 1-b_{i\,n}b_{n\,i}\,, 
\eq
with
\bq
b_{\alpha\,\beta}=-\left(\frac{11}{2}-\frac{8}{5}A_{\alpha\,\beta}^*-\frac{6}{5}B_{\alpha\,\beta}^*\right)\,\left(1+\frac{m_{\alpha}}{m_{\beta}}\right)^{-1}\,\frac{\nu_{\beta\,\alpha}}{\nu_{\alpha}^*}\,.
\eq
where the first bracket in the above expression is $2.7$ for $A_{\alpha\,\beta}^*=B_{\alpha\,\beta}^*=1$. The expression for $\chi_{n\Lambda}$ in the Eq.~(\ref{eq:tcn}) is taken from \cite{A64} as the expression given by \cite{Z02} contains a misprint.

As $\xiis \sim \xins \Delta_{\lambda} \sim 1$, the ion and neutral thermal conductivity given in Eq.~(\ref{eq:tci})-(\ref{eq:tcn}) becomes
\bq
\chi_{i\parallel}\approx\frac{5}{2}\frac{k_B}{m_i}\frac{p_i}{\nuis}\,, \quad \chi_{n\parallel}\approx\frac{5}{2}\frac{k_B}{m_n}\frac{p_n}{\nuns}\,.
\label{eq:tca}
\eq
The ion thermal conductivity is similar to Eq.~(4.40) of \cite{BR65} if we set $\nuis = \nu_{ii}$. However, due to the partially ionized nature of the solar plasma, ion–neutral collisions ($\nu_{in}$) dominate over ion–ion collisions ($\nu_{ii}$) throughout the entire chromosphere, i.e., $\nuis \approx \nu_{in}$. This dominance is reflected in Eqs.~(\ref{eq:tci}) and in (\ref{eq:tca}).

Braginskii’s parallel ion conductivity $\chi_{i\parallel}$ for fully ionized plasmas \citep{BR65} (which depends only on $\nu_{ii}$) has sometimes been extended to neutrals by substituting $p_n$ for $p_i$ and $\nu_{nn}$ for $\nu_{ii}$ \citep{S15, B19}. That extension assumes no coupling between plasma and neutral species, an assumption invalid in partially ionized media where neutral–ion/electron collisions strongly couple transport and ion–neutral drift induces ambipolar diffusion. Hence Braginskii-like formulas applied unchanged to partially ionized multicomponent plasmas are unreliable.

Both viscosity and thermal conductivity scale as $T^{1/2}$. From Eq.~(\ref{eq:tcn}), we observe that the neutral cross component is negligible, $\chi_{n\Lambda}\approx 0$, as $\Delta_{\lambda}\sim\xins\sim \mathcal{O}(1)$. In weakly ionized plasma, following \cite{SN09} and using Eq.~(\ref{eq:nus}) with $\nuns\approx 0.4\,\nu_{nn}$, we obtain
\begin{equation}
\chi_{n\parallel}\approx \frac{5}{4}\frac{k_B}{m_n}\frac{p_n}{\nuns}\,.
\end{equation}
For partially ionized conditions, this becomes $\nuns\approx 0.4\,\nu_{nn} (1+X_e)$. In summary, neutral thermal conductivity (both parallel and perpendicular components) dominates in the photosphere-chromosphere and in prominences/cold filaments within the corona, whereas the parallel and cross ion thermal components dominate in the transition region.

In Fig.~(\ref{fig:F5}), the dominant parallel and perpendicular conductivities are shown as functions of height. Due to the extremely weak dependence of neutral perpendicular conductivity on ion magnetization (i.e., on the ion Hall parameter, $\beta_i$), we find $\chi_{n\perp} \approx \chi_{n\parallel}$ throughout the photosphere-chromosphere [panels (a) and (b)]. Only in the transition region does $\chi_{i\parallel}$ dominate over $\chi_{n\parallel}$. Comparing panels (a) and (c), we conclude that the cross conductivity $\chi_{n\Lambda}$ is orders of magnitude smaller than both $\chi_{n\parallel}$ and $\chi_{n\perp}$. Furthermore, comparing panels (b) and (d), we infer that the cross conductivity is of similar order to the parallel ion conductivity in the transition region.

In summary, thermal conduction in the photosphere and chromosphere is dominated by neutrals, with ion thermal conductivity becoming important only in the transition region. The Braginskii model \citep{BR65}, derived for electron-ion plasmas, is therefore valid only for the transition region and corona—not throughout the entire solar atmosphere as has been assumed \citep{N22}.

\section{Wave Heating processes in the Solar Atmosphere}
Neutral-ion collisions in the weakly ionized solar atmosphere drive efficient wave damping across multiple spatial scales \citep{S15, S21, M25}. As demonstrated in \cite{PW24}, magnetic (Pedersen) diffusion and viscous (both parallel and perpendicular) momentum transport operate as wave damping mechanisms at distinct scale heights. Here, we demonstrate that thermal diffusion due to neutral-ion interactions provides an additional damping channel that can contribute significantly to chromospheric heating. We evaluate the relative importance of these wave heating mechanisms and compare their contributions to the overall energy budget.

Temperature fluctuations can be expressed in terms of density fluctuations. For an adiabatic process with equation of state $p\,\rho^{-\gamma} = \text{const}$, we have:
\bq
\frac{\delta T}{T_0} = \frac{\delta P}{P_0} - \frac{\delta \rho}{\rho_0} = (\gamma-1)\frac{\delta \rho}{\rho_0}\,.
\label{eq:dtf}
\eq
The first equality follows from the ideal gas law $P = \rho\, T$, while the second equality uses the adiabatic relation between pressure and density fluctuations. This adiabatic approximation is valid on timescales that are slow compared to the characteristic plasma timescales, over which dissipative effects can be neglected.

The entropy production rate $\Theta$ due to thermal conductivity is given by 
\bq
\Theta=\frac{1}{T_0^2}\left[\chi_{\parallel}\,\overline{\left(\nabla_{\parallel}T\right)^2}+\chi_{\perp}\,\overline{\left(\nabla_{\perp}T\right)^2}\right],
\eq
where $\chi=\chi_i + \chi_n$ represents the sum of ion and neutral thermal conductivity components. For temperature fluctuations, the right hand side of the above equation becomes
\bq
\left(\chi_{\parallel}\,k_{\parallel}^2 + \chi_{\perp}\,k_{\perp}^2\right)\Bigg[\left(\frac{\delta T}{T_0}\right)^2= \left(\gamma-1\right)^2\left(\frac{\delta \rho}{\rho_0}\right)^2\Bigg]\,.
\eq
Thus, the wave heating rate, which is related to the entropy production rate becomes \citep{BR65}
\bq
\Gamma_{\mbox{ther}}=\left(\gamma-1\right)^2\,\frac{k_{\perp}^2\,k_B\,T_0}{k^2\,\rho_0\,v_A^2}\left(\chi_{\parallel}\,k_{\parallel}^2 + \chi_{\perp}\,k_{\perp}^2\right)
\eq
Assuming $k_{\perp}=k\,\sin\theta\,,$\,$k_{\parallel}=k\,\cos\theta$, and $\chi_{\perp}=\chi_{\parallel}=\chi$  the above equation in terms of plasma beta, $\beta=2\,c_s^2/v_A^2$, which is a ratio of thermal and magnetic energies, with adiabaticity index $\gamma=5/3$ can be written as
\bq
\Gamma_{\mbox{ther}}=\frac{2}{9}\,\beta\,\frac{\chi\,k^2}{n_0}\sin^2\theta\,.
\eq

The cutoff wavenumber associated with perpendicular thermal conductivity,
\bq
k_{\chi}=\frac{n_0}{\chi}\,\va\,,
\eq
indicates that waves cannot propagate beyond a characteristic wavelength $\lambda_{\chi}=2\pi/k_{\chi}$ in the medium. This cutoff mechanism is analogous to those arising from viscous effects \citep{PW22},
\bq
k_{\nu}=\frac{\va}{\nu}\,,
\eq
and magnetic diffusion, namely due to Pedersen diffusion,
\bq
k_{\eta_P}=\frac{\va}{\eta_P}\,,
\eq
where each defines a characteristic scale below which the respective dissipative process dominates wave propagation.

\begin{figure}
\includegraphics[scale=0.25]{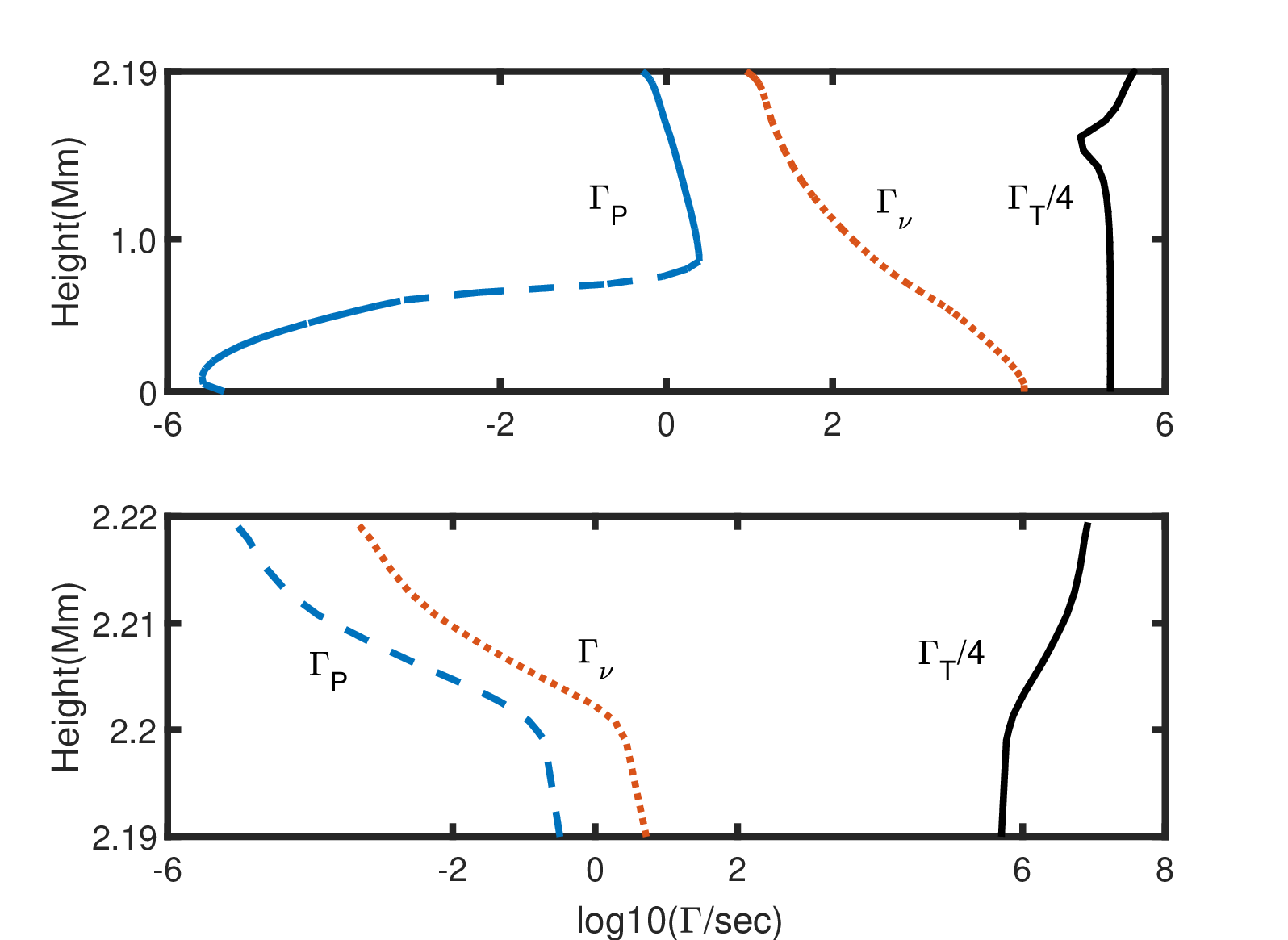}
\caption{The figure above shows the heating rates from wave damping through three mechanisms: thermal conductivity $\Gamma_T$ (solid line), viscosity $\Gamma_{\nu}$ (dotted line), and Pedersen diffusion $\Gamma_P$ (dashed line).
}
 \label{fig:F12}  
\end{figure}

The heating rate due to viscous damping of magnetosonic waves \citep{PW24}
\bq
\Gamma_{\nu}\approx \left(1+0.33\,\sin^2\theta\right)k^2\,\nu_0,,
\eq
(where a typo in their Eq.(29) has been corrected here), shown as the dotted curve in Fig.(\ref{fig:F12}), is much smaller than the thermal heating rate (solid curve). The damping due to Pedersen magnetic diffusion, $\Gamma_P=k^2\,\eta_P$ (dashed curve in the figure), is the smallest among these three non-ideal MHD effects.

\section{Summary}
The resonant interaction between the like charged and neutral pair is about an order of magnitude larger than the nonresonant interaction in the solar atmosphere.  The resulting ion mobility measured by ion-Hall beta $\beta_i$ is an order of magnitude smaller than the value calculated using only nonresonant collision frequencies. However, like previous works \citep{PW12, M12, PW13, K14}  Ohm diffusion dominates in the photosphere where $\eta_P=\eta_O$, while ambipolar diffusion is dominant in the upper chromosphere where $\eta_P=\eta_A$. Hall, $\eta_H$ which is sandwiched between the Ohm and ambipolar regimes, is important in the lower and middle
chromosphere.

In the partially ionized plasma, although viscosity and conductivity is due to both the ions and neutrals, the neutral
contribution is orders of magnitude larger than the ion contribution. This is because, unlike the fully ionized case when the perpendicular component is smaller than the parallel component by $1/\beta_i^2$, and becomes comparable only when $\beta_i\rightarrow 1$, the perpendicular neutral viscosity and thermal conductivity in a partially ionized case are almost independent of the ion magnetization. Therefore, the viscosity and thermal conductivity are mainly due to neutrals in the photosphere-chromosphere and lower transition region, and have both  parallel and perpendicular components. It is only closer to the coronal boundary that viscous momentum transport and thermal conduction is parallel to the magnetic field and is due to ions. The perpendicular component of the transport coefficients are negligible $\sim \mathcal{O}(1/\beta_i^2$) in this region.

Wave damping can very efficiently heat the ambient plasma. Among the main energy dissipation mechanisms, important under the solar conditions, are the Ohm and ambipolar diffusion, viscosity and thermal conductivity. 
As can be seen from Table 1, except for the transition region, it is the neutral component of the viscosity and thermal conductivity that dominates in the photosphere-chromosphere region of the sun. Since heating due to thermal conductivity $\Gamma_T$ is several orders of magnitude larger than viscous or Pedersen damping mechanisms, we conclude that thermal conductivity plays an important role in heating the chromospheric plasma.

Wave damping in a partially ionized solar atmosphere has been reviewed recently by \cite{B18}. While the damping mechanisms discussed in the present work are similar in nature to those described in that review, there are important differences. In particular, the collision frequencies, viscosities, and thermal conductivities adopted here differ from those used by \cite{B18}. Since the viscosity and thermal conductivity coefficients in \cite{B18} appear to be somewhat overestimated, this likely leads to stronger wave damping or shorter cutoff lengths \citep{S15}. A detailed quantitative comparison, however, lies beyond the scope of the present study.

We summarize in the following table the important transport coefficient in the various region of the solar atmosphere.

\begin{table*}
\small
\begin{flushleft}   
\begin{minipage}{110mm}
\caption{Transport Coefficients for F93 Model C for $B_0=100\,$ G footpoint field.}
\label{mathmode}
\begin{tabular}{|p{4cm}|p{2cm}|p{3cm}|p{3cm}|p{1.5cm}|}
\toprule
{Transport Coefficients}&  Photosphere  & lower and middle chromosphere      & upper chromosphere     & Transition region \\[2pt]    
\midrule
Magnetic Diffusivity &$\eta_O$             & $\eta_H$        & $\eta_A$     & $\eta_A$  \\[2pt]
\midrule
Viscosity & $\nu_{n0}\,,\nu_{n1}\,,\nu_{n2}$ & $\nu_{n0}\,,\nu_{n1}\,,\nu_{n2}$  &  $\nu_{n0}\,,\nu_{n1}\,,\nu_{n2}$ & $\nu_{i0}\,,\nu_{i3}$ \\[2pt]
\midrule
Thermal Conductivity & $\chi_{n\parallel}\,,\chi_{n\perp}$ & $\chi_{n\parallel}\,,\chi_{n\perp}$  &  $\chi_{n\parallel}\,,\chi_{n\perp}$ & $\chi_{i\parallel}\,,\chi_{i\Lambda}$ \\[2pt]
\bottomrule
\end{tabular}
\end{minipage}
\end{flushleft}
\end{table*}

\section*{Data Availability Statement}

Data sharing not applicable-no new data generated.

\end{document}